%% file: main.tex
\documentclass[sigconf,screen]{acmart}

\AtBeginDocument{%
  }

\setcopyright{cc}
\setcctype{by}
\acmDOI{10.1145/3832783.3834356}
\acmYear{2026}
\copyrightyear{2026}
\acmISBN{979-8-4007-2882-2/2026/10}
\acmConference[ASE '26]{Proceedings of the 41st IEEE/ACM International Conference on Automated Software Engineering}{October 12--16, 2026}{Munich, Germany}
\acmBooktitle{Proceedings of the 41st IEEE/ACM International Conference on Automated Software Engineering (ASE '26), October 12--16, 2026, Munich, Germany}
\acmSubmissionID{ase26main-p529-p}
\received{2026-03-26}
\received[accepted]{2026-06-18}

\usepackage[utf8]{inputenc}
\usepackage{microtype}
\usepackage{cleveref}
\usepackage{xspace}
\usepackage[ruled,vlined,linesnumbered]{algorithm2e}
\usepackage[shortlabels]{enumitem}
\usepackage{listings}
\usepackage{xcolor}
\usepackage{tikz}
\usepackage{subfigure}
\usepackage{graphicx}
\usepackage[most]{tcolorbox}
\usepackage{wrapfig}
\usepackage{romannum}
\usepackage{multirow}
\usepackage{pifont}

\input{sections/defs}

\begin{document}

% \pagenumbering{arabic}

\title{Automated Lemma Discovery in Agentic Program Verification}

\author{Huan Zhao}
\authornote{The first two authors contributed equally to this work and are listed in random order.}
\orcid{0009-0004-6412-0666}
\affiliation{%
  \institution{National University of Singapore}
  %\city{Singapore}
  \country{Singapore}
}
\email{zhaohuan@comp.nus.edu.sg}

\author{Haoxin Tu}
\authornotemark[1]
\authornote{Corresponding Author.}
\orcid{0000-0003-2389-1881}
\affiliation{%
  \institution{National University of Singapore}
  %\city{Singapore}
  \country{Singapore}
}
\email{haoxin.tu@nus.edu.sg}

\author{Zhengyao Liu}
\orcid{0009-0008-6062-5084}
\affiliation{%
  \institution{National University of Singapore}
  %\city{Singapore}
  \country{Singapore}
}
\email{zhengyao.liu@u.nus.edu}

\author{Martin C. Rinard}
\orcid{0000-0001-8095-8523}
\affiliation{%
  \institution{Massachusetts Institute of Technology}
  \city{Cambridge}
  \country{USA}
}
\email{rinard@csail.mit.edu}

\author{Abhik Roychoudhury}
\orcid{0000-0002-7127-1137}
\affiliation{%
  \institution{National University of Singapore}
  %\city{Singapore}
  \country{Singapore}
}
\email{abhik@nus.edu.sg}

\renewcommand{\shortauthors}{H. Zhao, H. Tu, Z. Liu, M. C. Rinard, and A. Roychoudhury}

%%
%% The abstract is a short summary of the work to be presented in the article.

\begin{abstract}
Deductive verification provides strong correctness guarantees for code by extracting verification conditions (VCs) and writing formal proofs for them.
The expertise-intensive task of VC proving is the main bottleneck in this process, and has been partly automated owing to recent advances in Large Language Model (LLM) agents.
However, existing proof agents are not able to discover helper lemmas -- auxiliary lemmas that aid in proving -- and thus fall short as programs grow in size and complexity. 

In this paper, we argue that VC proving for program verification is more than a purely mathematical task, and benefits considerably from program comprehension.
Our key insight is that human proof engineers often discover and apply helper lemmas based on their understanding of the program semantics, which are \emph{not} directly reflected in the VCs produced by VC generators.
Inspired by this insight, we propose an LLM agent, \toolname, that discovers helper lemmas in two ways.
Specifically, the agent first synthesizes lemmas \emph{offline} by directly analyzing the source code and specifications and then relating this semantic understanding to the mechanical, verbose encoding produced by VC generators. As the proof unfolds, \toolname then adapts existing helper lemmas \emph{online} to accommodate evolving proof states, enabling the agent to effectively discharge complex VCs on-the-fly.

We implement \toolname on top of an existing proof agent \autorocq for \rocq and the \framac ecosystem, and evaluate it on SV-COMP and established real-world subjects, including modules of the Linux kernel, Contiki OS, standard C++ library, and X.509 parser. Our experimental results demonstrate that \toolname significantly outperforms state-of-the-art approaches, highlighting the importance of program comprehension-aided lemma discovery in agentic program verification.
\end{abstract}






\settopmatter{printacmref=true} % Removes citation information below abstract
%\settopmatter{printfolios=true}
  
\maketitle

\input{sections/introduction}

\input{sections/background}

\input{sections/motivation}

\input{sections/approach}

\input{sections/evaluation}

\input{sections/discussion}

\input{sections/related}

\input{sections/conclusion}

\begin{acks}
We would like to sincerely thank all the anonymous reviewers for their valuable feedback and insights.
This research is supported by the National Research Foundation, Singapore, under its Artificial Intelligence (AI)-for-Science (AI4S) Challenge Grant (Award No. NRFAI4SCH-2025-0003), called ``AI for Program Reasoning''. Any opinions, findings and conclusions or recommendations expressed in this material are those of the authors and do not reflect the views of National Research Foundation.
\end{acks}

\input{sections/data-statement}

\bibliographystyle{ACM-Reference-Format}
\balance
\bibliography{refs}

% \appendix

\end{document}

%% file: sections/defs.tex
\newcommand\code[1]{{\tt\small #1}}
\definecolor{mygreen}{rgb}{0,0.5,0}
\definecolor{myblue}{rgb}{0,0,1}
\definecolor{mymauve}{rgb}{0.58,0,0.82}
\definecolor{myblack}{rgb}{0.24,0.17,0.12}
\definecolor{verylightgray}{rgb}{0.85,0.85,0.85}
\definecolor{dkgreen}{rgb}{0,0.6,0}
\definecolor{dkblue}{rgb}{0,0.4,0.4}
\definecolor{dkviolet}{rgb}{0.3,0,0.5}
\definecolor{dkred}{rgb}{0.6,0,0}
\definecolor{dkpink}{rgb}{0.9,0.3,0.5}
\definecolor{dkyellow}{rgb}{0.8,0.8,0}

\lstdefinelanguage{Coq}{ 
    mathescape=true,
    texcl=false, 
    morekeywords=[1]{Section, Module, End, Require, Import, Export,
        Variable, Variables, Parameter, Parameters, Axiom, Hypothesis,
        Hypotheses, Notation, Local, Tactic, Reserved, Scope, Open, Close,
        Bind, Delimit, Definition, Let, Ltac, Fixpoint, CoFixpoint, Add,
        Morphism, Relation, Implicit, Arguments, Unset, Contextual,
        Strict, Prenex, Implicits, Inductive, CoInductive, Record,
        Structure, Canonical, Coercion, Context, Class, Global, Instance,
        Program, Infix, Theorem, Lemma, Corollary, Proposition, Fact,
        Remark, Example, Proof, Goal, Save, Qed, Defined, Hint, Resolve,
        Rewrite, View, Search, Show, Print, Printing, All, Eval, Check,
        Projections, inside, outside, Def},
    morekeywords=[2]{forall, exists, exists2, fun, fix, cofix, struct,
        match, with, end, as, in, return, let, if, is, then, else, for, of,
        nosimpl, when},
    morekeywords=[3]{Type, Prop, Set, true, false, option},
    morekeywords=[4]{pose, set, move, case, elim, apply, clear, hnf,
        intro, intros, generalize, rename, pattern, after, destruct,
        induction, using, refine, inversion, injection, rewrite, congr,
        unlock, compute, ring, field, fourier, replace, fold, unfold,
        change, cutrewrite, simpl, have, suff, wlog, suffices, without,
        loss, nat_norm, assert, cut, trivial, revert, bool_congr, nat_congr,
        symmetry, transitivity, auto, split, left, right, autorewrite},
    morekeywords=[5]{by, done, exact, reflexivity, tauto, romega, omega,
        assumption, solve, contradiction, discriminate},
    morekeywords=[6]{do, last, first, try, idtac, repeat},
    morekeywords=[7]{shift, offset, region, base, valid_rw, valid_rd},
    morecomment=[s]{(*}{*)},
    showstringspaces=false,
    morestring=[b]",
    morestring=[d]’,
    tabsize=3,
    extendedchars=false,
    sensitive=true,
    breaklines=false,
    basicstyle=\footnotesize,
    captionpos=b,
    columns=[l]flexible,
    identifierstyle={\ttfamily\color{black}},
    keywordstyle=[1]{\bf\ttfamily\color{black}},
    keywordstyle=[2]{\ttfamily\color{dkviolet}},
    keywordstyle=[3]{\ttfamily\color{dkblue}},
    keywordstyle=[4]{\ttfamily\color{dkblue}},
    keywordstyle=[5]{\ttfamily\color{dkblue}},
    keywordstyle=[6]{\ttfamily\color{dkred}},
    keywordstyle=[7]{\ttfamily\color{black}},
    stringstyle=\ttfamily,
    commentstyle={\ttfamily\color{dkgreen}},
    literate=
    {\\forall}{{\color{dkgreen}{$\forall\;$}}}1
    {\\exists}{{$\exists\;$}}1
    {<-}{{$\leftarrow\;$}}1
    {=>}{{$\Rightarrow\;$}}1
    {==}{{\code{==}\;}}1
    {==>}{{\code{==>}\;}}1
    {->}{{$\rightarrow\;$}}1
    {<->}{{$\leftrightarrow\;$}}1
    {<==}{{$\leq\;$}}1
    {\#}{{$^\star$}}1 
    {\\o}{{$\circ\;$}}1 
    {\@}{{$\cdot$}}1 
    {\/\\}{{$\wedge\;$}}1
    {\\\/}{{$\vee\;$}}1
    {++}{{\code{++}}}1
    {~}{{$\sim$}}1
    {\@\@}{{$@$}}1
    {\\mapsto}{{$\mapsto\;$}}1
    {\\hline}{{\rule{\linewidth}{0.5pt}}}1
}[keywords,comments,strings]

\newcommand{\tightcolorbox}[2]{%
  \begingroup
  \setlength{\fboxsep}{0pt}% no padding
  \colorbox{#1}{#2}%
  \endgroup
}

\newtcolorbox{promptresponsebox}[1][]{
  colback=white,               % remove background color (set to white)
  colframe=black,              % frame color (your defined dark blue)
  fonttitle=\bfseries, 
  coltitle=black,              % make title font black
  colbacktitle=gray!20,        % light gray background for title
  boxrule=0.5pt,               % thin border
  top=1pt,                     % reduce top padding
  bottom=1pt,                  % reduce bottom padding
  left=4pt,                    % reduce left padding
  right=4pt,                   % reduce right padding
  toptitle=0.5pt,                % reduce title top margin
  bottomtitle=0.5pt,             % reduce title bottom margin
  #1
}

\newtcolorbox{resultbox}{
  colback=gray!10,     % background color
  colframe=gray!70,    % border color
  boxrule=0.8pt,       % border thickness
  arc=4pt,             % rounded corners
  left=4pt, right=4pt, top=4pt, bottom=4pt, % padding
  enhanced
}

\theoremstyle{definition}
\newtheorem*{definition*}{Definition}
\usetikzlibrary{arrows.meta,positioning,shapes.geometric}

\newcommand{\eg}{e.g.,\xspace}
\newcommand{\ie}{i.e.,\xspace}
\newcommand{\rocq}{\textsc{Rocq}\xspace}

\newcommand{\framac}{\textsc{Frama-C}\xspace}
\newcommand{\hammer}{\textsc{CoqHammer}\xspace}
\newcommand{\copra}{\textsc{Copra}\xspace}

\newcommand{\autorocq}{\textsc{AutoRocq}\xspace}
\newcommand{\toolname}{\textsc{LemmaNet}\xspace}
\newcommand{\varNoOnline}{\text{\toolname($\neg$Onl)}\xspace}
\newcommand{\varNoOffline}{\text{\toolname($\neg$Ofl)}\xspace}
\newcommand{\varNaiveOffline}{\text{\toolname($\neg$PSA)}\xspace}

\newcommand{\dafny}{\textsc{Dafny}\xspace}
\newcommand{\verus}{\textsc{Verus}\xspace}
\newcommand{\viper}{\textsc{Viper}\xspace}

\makeatletter
\let\originalformatdoi\@formatdoi
\renewcommand{\@formatdoi}[1]{%
  \hspace{0.1em}\originalformatdoi{#1}%
}
\makeatother

%% file: sections/introduction.tex
\section{Introduction}

%\paragraph{Need of formal verification in the age of coding agents}
Trusted software is critical for the safety and security of modern computing systems, especially in safety-critical domains such as aerospace, automotive, and healthcare \cite{trusted-roadmap,leroy2016compcert,DBLP:conf/sosp/KleinEHACDEEKNSTW09}. 
In pursuit of this goal, formal verification is a powerful technique that can provide strong guarantees that the software correctly implements its specification~\cite{seligman2023formal,DBLP:conf/lpar/Leino10,verified_db}. 
One widely used approach is {\em deductive verification}, which involves generating mathematical verification conditions (VCs) from source code and specifications, then using interactive theorem provers to generate formal machine-checked proofs of these VCs~\cite{almeida2010deductive}. Proving these VCs is currently an extremely labor- and expertise-intensive undertaking, which has significantly hindered its adoption in practice \cite{DBLP:conf/sosp/KleinEHACDEEKNSTW09, leroy2016compcert}. 
With recent advances in the 
mathematical reasoning ability of large language models (LLMs)~\cite{alon2025integrating,alpha-proof}, there has been growing interest in leveraging LLM agents to assist in the formal verification process \cite{lu2025palm,copra24,autorocq}. 

\noindent\paragraph{Limitations of Existing Approaches.}
Existing theorem-proving approaches formulate VC proving as a purely mathematical exercise \cite{sanchez2023passport,sanchez2020generating,lu2025palm}, where the input to the system is a theorem statement, and the output is a sequence of proof steps that establish its correctness. Recent approaches that adopt this paradigm have therefore focused on improving various aspects of this process~\cite{autorocq,copra24,rango-icse25,sanchez2024qedcartographer}. 
Human proof engineers take a different approach. When discharging complex obligations, they typically start with an understanding of the program semantics and propose helper lemmas that capture relevant aspects of the program structure and properties~\cite{dodds2013mostly, nguyen2008enhancing}. These helper lemmas often identify key observations and/or intermediate results that can considerably simplify the proving process, with the helper lemmas often refined and enhanced in light of new information obtained as the proof progresses. 
This aspect of proof development remains largely overlooked by existing approaches.
As such, they struggle to discharge complex VCs that benefit from insights into the program semantics, structure, and properties.

\paragraph{New Angle to VC Proving.}
We argue that program VC proving is \emph{not} purely a matter of mathematical reasoning, but should be deeply coupled with program comprehension. 
Our argument is based on the fact that proofs are easier to develop and interpret when they align with program structure and source-level semantics, with the alignment driven by intermediate lemmas that capture important insights into the program structure, semantics, and properties~\cite{dodds2013mostly, nguyen2008enhancing}.
%\zh{I feel referring to figures on future pages makes it really hard to read.}
For example, to prove a loop invariant obligation (\eg line 10 in \autoref{fig:hex2bin}), a human proof engineer usually first tries to understand the loop's behavior and write a proper semantics-aware VC that is amenable to straightforward proof (\eg the semantics-aware VC in Figure~\ref{fig:ghost_vc}).
This informal reasoning process is crucial for effectively discharging complex VCs and is often guided by the program semantics, which are not directly reflected in the proof-targeted VCs (\eg the proof-targeted VC in Figure~\ref{fig:original_vc}) generated by existing VC generators (\eg \framac).
% which tend to be verbose and have little to no connection to the program semantics, posing significant challenges to human or advanced LLMs to comprehend.
We therefore propose to mimic this human proof process by developing an LLM agent that discovers helper lemmas informed by program structure and semantics, aiming to bridge the gap between the program semantics and the proof-targeted VCs, and thus facilitate the proving process.

\paragraph{Challenges.}
Two challenges need to be addressed to enable such a human-like proving process for effective automated verification. %\todo{See if we like this better, this is option 1}
First, existing VC generators model the full complexity of the language semantics, including aspects such as integer overflow, pointer arithmetic, and memory addressing. The generated VCs obscure the conceptual structure of the program semantics in low-level details that generators use to model these and other aspects of the underlying language semantics. While a naive LLM-based approach can more clearly reflect the program structure, the approach is fundamentally unsound, and the VCs come with no correctness guarantees.  
%\todo{And here is option 2}
%First, constructing a semantics-aware VC that faithfully captures the behavior of an arbitrary input program is inherently difficult, whereas a naive LLM-based approach is fundamentally \emph{unsound}.
%This requires us to introduce formal guarantees into an otherwise informal reasoning process over program semantics without limiting its applicability.
%\todo{end of option 2}
%{haoxin: option 1 looks clearer and better! Thanks!}
Second, as VC complexity grows, it becomes difficult to foresee \emph{a priori} how proof states will evolve throughout the proving process. As such, proof strategies must dynamically adapt as new information is updated during proving while still preserving the coherence and relevance of the existing reasoning trajectory.

% Therefore, we will need to bridge the gap between the semantics-aware VC and the verbose proof-targeted VC generated by VC generators, so that the proof of the proof-targeted VC benefits from the proof of the semantics-aware VC, while ultimately discharging the VC produced by a trusted VC generator to ensure soundness.

% The capability to adapt proof strategies on-the-fly is crucial for effectively discharging complex VCs, as it allows the agent to iteratively refine its approach based on feedback from the proof process, which is often necessary for handling the intricacies of real-world programs.

\paragraph{Our Solution}
We propose a new solution, \toolname, that 
addresses the above two challenges through a carefully designed mechanism of {\em helper lemma discovery}. \toolname discovers 
helper lemmas in two stages of the verification workflow. 
First, \toolname performs \emph{offline} analysis before theorem proving starts to extract a semantics-aware VC directly from the annotated source code. To mitigate the potential untrustworthiness of LLM-based analysis, the semantics-aware VC is used merely as a reference for synthesizing helper lemmas to bridge source-level concepts with the proof-targeted VC. These helper lemmas are ultimately used to discharge the VC produced by a trusted VC generator, ensuring soundness throughout the process.
Second, \toolname performs \emph{online} lemma adaptation as proving progresses. This adaptation is achieved by iteratively refining and enhancing the helper lemmas based on the feedback from the theorem prover.
% by either strengthening the lemma statement or revising conflicting definitions that the lemma depends on through feedback-guided lemma adaptation.
Our results indicate that this online adaptation process often enables the agent to effectively discharge complex, otherwise intractable VCs while maintaining alignment with the program semantics captured by the semantics-aware VC (c.f. \autoref{sec:evaluation:understanding_helper_lemmas}).
By integrating both offline lemma synthesis and online lemma adaptation, \toolname is able to effectively discover and apply helper lemmas that are crucial for VC proving, significantly improving its efficacy in verifying real-world software projects.

\paragraph{Evaluation.}
Our evaluation on SVCOMP~\cite{svcomp} and NTP4VC~\cite{xu2026neural} benchmarks (which include Linux kernel modules, Contiki OS, X.509 parser, etc.) shows that \toolname can effectively discharge complex VCs arising from real-world verification tasks, significantly outperforming state-of-the-art proof agents \autorocq~\cite{autorocq} and \copra~\cite{copra24} by 26.8\%--51.7\%.
We also report what helper lemmas are proposed and how they aid in agentic program verification.

\paragraph{Contributions}
This paper makes the following contributions:
\begin{itemize}[leftmargin=1em,nosep]
    \item It proposes a novel agentic approach for lemma discovery to facilitate program verification.
    \item It instantiates this framework with \toolname\footnote{The tool is released at \url{https://github.com/NUS-Program-Verification/LemmaNet}.}, which supports both offline lemma synthesis and online lemma adaptation to effectively discharge complex VCs.
    \item It presents results from an evaluation of \toolname on leading benchmarks. These results highlight the effectiveness of \toolname in enabling the verification of complex, real-world systems such as Linux kernel modules.
\end{itemize}

\paragraph{Paper Organization} \autoref{sec:background} introduces the background on deductive verification and automated theorem proving. \autoref{sec:motivating-example} presents a motivating example that highlights the challenges of existing approaches and motivates our design. \autoref{sec:approach} describes the design and implementation of \toolname. \autoref{sec:eva} evaluates \toolname on real-world verification benchmarks and analyzes the efficacy of its key components. \autoref{sec:discussions} discusses limitations and potential future directions. \autoref{sec:related} reviews related work, and \autoref{sec:conclusion} concludes the paper.

%% file: sections/background.tex
\section{Background} \label{sec:background}

\subsection{Deductive Verification Practices}

Deductive verification formally checks if the code conforms to its formal specification. 
In contrast to testing, which can prove the \emph{existence} of bugs, deductive verification can prove their \emph{absence} by reasoning over all possible executions under a well-defined semantic model. 
The workflow starts with augmenting the source code with specification annotations either manually or automatically~\cite{specgen-icse25}.
These annotations are formal contracts that are attached to functions and program locations, and specify key properties such as pre/post-conditions, invariants, or (partial or full) correctness. \autoref{fig:hex2bin} presents example source code with ACSL~\cite{DBLP:journals/fac/KirchnerKPSY15} annotations.
The annotated source code is then transformed into a collection of formal \emph{verification conditions} (VCs)~\cite{efremov2018deductive,almeida2010deductive}, also known as proof obligations, via the use of a VC generator that applies a weakest precondition (or strongest postcondition) calculus to systematically produce logical formulas.
The validity of all VCs implies that the program meets its specification. 
This style of verification has been widely adopted in mature verification frameworks including \dafny~\cite{DBLP:conf/lpar/Leino10, DBLP:journals/pacmse/MisuLM024}, \verus~\cite{DBLP:journals/corr/abs-2303-05491}, \viper~\cite{DBLP:conf/vmcai/0001SS16}, and \framac~\cite{DBLP:journals/fac/KirchnerKPSY15}.

While some VCs can be discharged by automated solvers such as SMT solvers \cite{de2008z3,barbosa2022cvc5}, Interactive Theorem Provers (ITPs) are required for more complex obligations that involve sophisticated reasoning~\cite{nipkow2002isabelle,de2015lean,coq1996coq}.
The \rocq proof assistant (formerly \textsc{Coq})~\cite{coq1996coq} is a prominent example of ITPs. To discharge a VC as the proof \emph{goal} in \rocq, one applies a sequence of \emph{tactics}: commands that transform the current goal into simpler sub-goals. Each successful tactic application refines a \emph{proof term} that \rocq's kernel type-checks for soundness. 
The proof concludes when no sub-goal remains, yielding a certified derivation. 
To enable proofs of complex obligations, human provers often introduce auxiliary \emph{lemmas} that encapsulate reusable intermediate facts, which is a well-supported feature in modern ITPs such as \rocq.

\subsection{Theorem Proving for Verification}

\noindent\paragraph{Machine Learning for Proof Automation.}
Automating proof construction is a key step toward closing the verification gap. 
Recent advances in machine learning have spurred numerous approaches to this problem. 
Pioneering efforts~\cite{yang2019coqgym, sanchez2020generating} frame proof synthesis as sequence prediction, employing neural networks trained on syntactic patterns~\cite{yang2019coqgym}, proof-state encoding~\cite{first2020tactok}, and contextual features~\cite{sanchez2023passport}. 
These prediction models later became the cornerstone of more sophisticated approaches that employ various search algorithms to steer tactic selection toward genuine progress~\cite{blaauwbroek2020tactician, sanchez2024qedcartographer}.

Large Language Models (LLMs) offer strong mathematical reasoning capabilities, making them attractive for proof automation~\cite{alpha-proof}. 
Recent approaches have leveraged LLMs for whole-proof generation~\cite{lu2025palm} or retrieval-augmented tactic generation (RAG) using a curated database of historical proof states~\cite{rango-icse25}.
Notably, LLM agents show great promise in proof generation. \copra~\cite{copra24} iteratively proposes tactics, executes them, and incorporates feedback from the proof assistant to refine subsequent suggestions. \autorocq~\cite{autorocq} allows for a more dynamic workflow, enabling the agent to autonomously gather additional context or traverse the proof tree. 

\noindent\paragraph{Proof Automation for VCs.}
Despite notable progress, transferring these techniques to program-derived theorems remains difficult. 
Existing neural theorem provers~\cite{sanchez2020generating,lu2025palm,rango-icse25,copra24} are trained and evaluated primarily on mathematical lemmas with human-written ground truth~\cite{yang2019coqgym}. 
Prior work~\cite{autorocq} has shown that verification conditions arising from real-world programs tend to be more complex than their mathematical counterparts. %reflecting the structural and semantic intricacies of the underlying code. 
This complexity compounds as programs grow in scale, increasingly straining existing approaches for automated discharge of verification conditions.

A critical bottleneck lies in \emph{helper lemma discovery} \cite{kurashige2024cclemma,yang2019lemma,ta2017automated}. 
Complex VCs often cannot be proved directly; instead, they require auxiliary lemmas that capture intermediate facts about program semantics, data structure properties, or arithmetic relationships. 
%These helper lemmas bridge the gap between low-level program operations and high-level specification requirements. 
Existing approaches largely overlook this challenge, focusing on tactic generation and lemma retrieval from existing libraries only. 
However, in the context of VC proving, key lemmas are often \emph{not} available in the proof context. In practice, identifying and synthesizing the right helper lemmas demands a nuanced understanding of both the program under verification and the proof structure.

%% file: sections/motivation.tex
\section{Motivating Example} \label{sec:motivating-example}

In this section, we illustrate the challenges in the existing VC proving workflow through a motivating example, and present the key insights that motivate our solution. 

\noindent\paragraph{Illustration of a Proof-targeted VC}
\autoref{fig:hex2bin} shows a simplified C program from \texttt{\small hex2bin.c} in the Linux kernel codebase, which converts a string representing a hexadecimal number to its binary representation. Guarded by {\color{dkgreen}\texttt{//@}} and {\color{dkgreen}\texttt{/*@...*/}} are ACSL~\cite{DBLP:journals/fac/KirchnerKPSY15} annotations that declare the specifications to be proved. 
\tightcolorbox{verylightgray}{Highlighted} is a key loop invariant, whose validity underpins the verification of the program.
Specifically, it states an invariant that variable \texttt{\small src} is always at least as large as \texttt{\small osrc} throughout the \texttt{\small while} loop, where \texttt{\small osrc} is a ghost variable recording the original value of \texttt{\small src}.
Intuitively, the preservation of this simple invariant holds almost trivially, as \texttt{\small src} is monotonically incremented exactly by two in each iteration.

% 1. representation issue: little connection between the VC and the program semantics;
% 2. without proper representation, proposing bridge lemmas online is difficult/impractical (based on our experiments).
When passed through a VC generator such as \framac's WP plugin~\cite{wp}, however, this simple invariant yields a surprisingly complex VC as shown in Figure~\ref{fig:original_vc}.
We denote it as the {\it proof-targeted VC} $\phi_C$.
This proof-targeted VC, bearing little resemblance to the original program or the annotation, is a sprawling, low-level formula that encodes tool-specific details such as \framac's internal memory model for C, pointer arithmetic, and integer overflow semantics. 
Consequently, even state-of-the-art theorem proving agents such as \cite{autorocq} struggle to synthesize a proof for $\phi_C$.

\input{figures/motivating-example-src.tex}

\input{figures/motivating-example-lemmas.tex}

\noindent\paragraph{Illustration of a Semantics-aware VC}
If one formally expresses the high-level understanding that \texttt{\small src} is monotonically increased by two in each iteration, a proof engineer may produce a VC as shown in Figure~\ref{fig:ghost_vc}.
The lemma at line 6, denoted as the {\it semantics-aware VC} $\phi_A$, states that, for all \texttt{\small osrc} and \texttt{\small src}, $\texttt{\small osrc} \leq_\texttt{addr} \texttt{\small src} \to \texttt{\small osrc} \leq_\texttt{addr} \texttt{\small src}+2$. Here, the notion $\leq_\texttt{addr}$ refers to a custom, intuitive model of address comparison as defined in line 2.
Compared to Figure~\ref{fig:original_vc}, $\phi_A$ encodes and proves the same loop invariant but is instead expressed directly in terms of source-level concepts drawn from the program and its specification, making it much more compact and amenable to straightforward proof.
Indeed, $\phi_A$ could be easily proved by existing approaches, or by simply consulting an LLM.

\noindent\paragraph{Key Insights.}
Our key observation is that the inefficiency of existing VC proving approaches stems \emph{not} from lagging mathematical reasoning capability, but from the burial of semantic signal under layers of verbose, tool-specific encoding that has no direct correspondence to how the program is understood. 
As such, approaches that treat VC proving as a purely mathematical endeavor and reason directly about $\phi_C$ struggle even on simple programs and properties such as the one in \autoref{fig:hex2bin}.
This observation is consistent with the fact that human proof engineers often discover and apply helper lemmas that are informed by their understanding of the program semantics \cite{dodds2013mostly, nguyen2008enhancing}.
On the other hand, reasoning exclusively about the semantics-aware VC $\phi_A$ is \emph{insufficient} for soundness reasons, as there is no formal argument that $\phi_A$'s modeling of C constructs is faithful. 
In other words, the verifier must therefore ultimately discharge $\phi_C$ produced by the (trusted) VC generator, but without being overwhelmed by its low-level encoding. 
This calls for a verification framework that can leverage the insights from $\phi_A$ to facilitate the proof of $\phi_C$, while ultimately targeting the final proof towards $\phi_C$ to guarantee soundness.

%The structure of two-layered reasoning is reminiscent of the abstract data type (ADT) tradition~\cite{jones1983tentative}, where an abstract interface and a concrete implementation are reasoned about separately, e.g., the behaviors of a stack and its array-based instantiation.
%The two layers are then related via a \emph{retrieve function} that formally witnesses their correspondence. 
%Inspired by this tradition, we ask: can we similarly relate two layers of encoding through a formal bridge, so that the proof of $\phi_C$ benefits from the proof of $\phi_A$?

\noindent\paragraph{Two Challenges.}
Two challenges need to be addressed to realize the above vision.
First, the gap between $\phi_A$ and $\phi_C$ is non-trivial, and it is not clear how to systematically derive helper lemmas that can bridge the gap between them.
Second, as the proof state evolves during the proving process, the applicability of the helper lemmas may change. It is thus difficult to derive the exact helper lemmas that are useful to the proving process, calling for a dynamic approach that adapts the helper lemmas as proof states change.

\noindent\paragraph{Our Solution.}
% 1. offline helper lemma generation to bridge the gap between the program semantics and the VCs
% 2. oneline refinement to refine the helper lemmas on the fly in response to evolving proof states.
We introduce mechanisms to discover helper lemmas in two stages.
First, before the VC proving process starts, an \emph{offline} synthesizer discovers lemmas to bridge the gap between $\phi_A$ and $\phi_C$. This is achieved by analyzing the annotated program to derive $\phi_A$ and then generating lemmas that are aligned with both $\phi_A$ and $\phi_C$. These lemmas can be used to discharge $\phi_C$ while remaining faithful to the program semantics captured by $\phi_A$. 
Second, as the proof unfolds, an \emph{online} adapter refines these lemmas if they are inapplicable to the proof of $\phi_C$. 
The adaptation is based on real-time feedback from the proof assistant, and typically involves strengthening the lemma statement or revising conflicting representations that the lemma depends on.
As a result, the agent can effectively discharge complex VCs that are otherwise intractable, while maintaining the alignment with the program semantics captured by $\phi_A$.
As we can see in Figure~\ref{fig:helper_lemma}, the offline synthesizer discovers two important lemmas (\ie \xspace {\tt \small HL1\_addr\_le\_shift\_same\_base} and {\tt \small HL2\_addr\_le\_same\_base}), with which $\phi_C$ (in Figure \ref{fig:original_vc}) can be easily discharged.

%% file: figures/motivating-example-src.tex
\begin{figure}[t]
\centering
\begin{lstlisting}[escapechar=&,xleftmargin=0pt,xrightmargin=0pt,linewidth=0.98\linewidth]
int hex_to_bin(char ch) {...}

/*@ requires ...; ensures ... */
int hex2bin(u8 *dst, const char *src, size_t count)
{
	//@ ghost size_t ocount = count; 
    //@ ghost char *osrc = src;
	/*@ 
        loop invariant 0 <= count <= ocount;
        &\tightcolorbox{verylightgray}{\color{dkgreen}loop invariant osrc <= src;}&
	    loop invariant \forall char *p; ...
	 */
	while (count--) {
		int hi = hex_to_bin(*src++); 
        int lo = hex_to_bin(*src++);
		if ((hi < 0) || (lo < 0)) 
            return -1;
		//@ assert 0 <= ((hi << 4) | lo) <= 255;
		*dst++ = (hi << 4) | lo;
	}
	//@ assert count == ((size_t)-1);
	return 0;
}
\end{lstlisting}
%\vspace{-1em}
\caption{
Source code of \texttt{hex2bin.c} from \cite{efremov2018deductive} with ACSL {\color{dkgreen} annotation}. The loop invariant to prove is \tightcolorbox{verylightgray}{highlighted}.
}
\label{fig:hex2bin}
\end{figure}

%% file: figures/motivating-example-lemmas.tex
\begin{figure*}[t]
\centering
\begin{minipage}[c]{0.51\linewidth}
\centering
\subfigure[Proof-targeted VC $\phi_C$ in \rocq, generated by \framac (simplified) and the complete proof generated by \toolname. Helper lemmas are applied in lines 33-34.]{%
\parbox{\linewidth}{%
\begin{lstlisting}[language=Coq]
Definition is_sint32 (x:Numbers.BinNums.Z) : Prop := ...

(* Proof-targeted Verificaion Condition *)
Theorem wp_goal :
  forall (t:Z -> Z) (t1:addr -> Z) (t2:Z -> Z) 
  (a:addr) (i:Z) (i1:Z) (i2:Z) (a1:addr) (a2:addr),
  let x := (16 * i) in let x1 := lor i1 x in 
  let x2 := offset a2 in let x3 := offset a1 in 
  let x4 := t1 a2 in let x5 := t2 (to_uint8 x4) in 
  let x6 := t1 (shift a2 1) in let x7 := t2 (to_uint8 x6) in 
  let x8 := land 68 x5 in let x9 := land 68 x7 in
  ~ (i2 = 0) -> (x = (lsl i 4)) -> ((i1 + x) = x1) ->
  ((Z.quot (x2 + ((-1) * x3)) 2) = 0) -> 
  ((Z.rem (x3 + ((-1) * x2)) 2) = 0) ->
  (0 <= i2) -> (0 <= i1) -> (0 <= i) -> 
  ((region (base a1)) <= 0) -> ((region (base a)) <= 0) ->
  (0 <= x1) -> ((to_uint64 ((-1) + i2)) <= i2) -> 
  (i1 <= 15) -> (i <= 15) -> 
  (x1 <= 255) -> IsArray_uint8 t2 -> linked t -> sconst t1 -> 
  is_sint32 i1 -> is_sint32 i -> is_uint64 i2 -> addr_le a1 a2 -> 
  addr_le a1 a1 -> addr_le a a -> is_sint8 x4 -> is_uint8 x5 -> 
  is_sint8 x6 -> valid_rw t (shift a 0) (1 + i2) -> 
  valid_rd t (shift a1 0) (1 + (2 * i2)) -> is_uint8 x7 -> 
  ((x8 = 0) -> (i = (-1))) -> (~(x8 = 0) -> ((L_hex_to_bin x4) = i)) ->
  ((x9 = 0) -> (i1 = (-1))) -> (~(x9 = 0) -> ((L_hex_to_bin x6) = i1)) ->
  (forall (a3:addr), addr_lt a3 a2 -> addr_le a1 a3 -> 
  ~ ((land 68 (t2 (to_uint8 (t1 a3)))) = 0)) ->
  (forall (a3:addr), addr_lt a3 a1 -> addr_le a1 a3 -> 
  ~ ((land 68 (t2 (to_uint8 (t1 a3)))) = 0)) ->
  addr_le a1 (shift a2 2).
Proof.
  intros t t1 i t2 a i1 i2 a1 a2; cbn. ...
  apply (HL1_addr_le_shift_same_base a1 a2 2).
  apply (HL2_addr_le_same_base _ _ Hle_a1a2).
  all: try (exact Hle_a1a2); try lia.
Qed.
\end{lstlisting}%
}%
\label{fig:original_vc}
}
\end{minipage}
\hfill
\begin{minipage}[c]{0.47\linewidth}
\subfigure[Semantics-aware VC $\phi_A$ expressed and proved in source-level concepts.]{%
\parbox{\linewidth}{%
\begin{lstlisting}[language=Coq]
(* Custom model of address comparison *)
Definition addr_le_model (p q:addr) : Prop :=
  base p = base q /\ offset p <= offset q.

(* Semantics-aware Verificaion Condition *)
Lemma goal_osrc_le_src_preserved :
  forall osrc src : addr,addr_le_model osrc src ->
    addr_le_model osrc (shift src 2).
Proof.
  intros osrc src [Hb Hoff]. split.
  (* base preserved *)
  - unfold shift, base in *; simpl in *; exact Hb. 
  (* offset incremented *)
  - unfold shift, offset in *; simpl in *; lia.
Qed.
\end{lstlisting}%
}%
\label{fig:ghost_vc}
}
\subfigure[Helper lemmas discovered by \toolname, which are used to bridge the gap between semantics-aware and proof-targeted VCs.]{%
\parbox{\linewidth}{%
\begin{lstlisting}[language=Coq]
Lemma base_shift :
  forall p k, base (shift p k) = base p. Proof... Qed.

Lemma offset_shift : 
  forall p k, offset (shift p k) = (offset p + k)%Z. Proof... Qed.

(* Helper Lemma 1 *)
Lemma HL1_addr_le_shift_same_base :
  forall p q k, base p = base q -> addr_le p q
   -> (0 <= k)%Z -> addr_le p (shift q k).
Proof. 
  rewrite base_shift. rewrite offset_shift... (* steps omitted *)
Qed.

(* Helper Lemma 2 *)
Lemma HL2_addr_le_same_base : 
  forall p q, addr_le p q -> base p = base q.  Proof... Qed.
\end{lstlisting}%
}%
\label{fig:helper_lemma}
}
\end{minipage}
%\vspace{-1em}
\caption{(a) Proof-targeted VC $\phi_C$ v/s (b) semantics-aware VC $\phi_A$ for the highlighted loop invariant in \autoref{fig:hex2bin}. (c) Helper lemmas are discovered by \toolname, and are useful in generating a complete proof for the proof-targeted verification condition. }
\label{fig:vc-cmp}
\end{figure*}

%% file: sections/approach.tex
\section{The Design of \toolname} \label{sec:approach}

\subsection{System Overview}  
\label{sec:approach:system_overview}

\autoref{fig:system} presents an overview of \toolname.
Built on top of an existing tactic-by-tactic proof agent that directly reasons about the proof-targeted VC (\ding{192}$\to$\ding{193}$\to$\ding{194}$\to$\ding{195}), \toolname's novelties lie in two new components: (1) an \emph{offline lemma synthesizer} that synthesizes helper lemmas that are aligned with the program semantics before the prover is invoked (\autoref{sec:approach:offline_discovery}, \ding{196}), and (2) an \emph{online lemma adapter} that iteratively refines the helper lemmas and their supporting obligations to accommodate evolving proof states during the proving process (\autoref{sec:approach:online_refinement}, \ding{197}).
Concretely, the offline lemma synthesizer extracts program semantics from the source code and specification via a \emph{program semantic analyzer} (\autoref{sec:approach:offline_discovery:psa}), and generates helper lemmas that are aligned with both the semantics-aware VC (derived from program semantics) and the proof-targeted VC (derived from the verification tool).
While the bridging may fail due to mismatches between the semantics-aware VC and the proof-targeted VC, the online lemma adapter maintains a dynamic library of helper lemmas via an \emph{adaptive lemma maintainer} (\autoref{sec:approach:alm}), and refines the lemmas on-the-fly in response to evolving proof states through \emph{feedback-guided lemma adaptation} (\autoref{sec:approach:fla}).
The two components work in tandem to enable powerful reasoning and adaptability, enabling the agent to effectively discharge complex, otherwise intractable VCs.

\input{figures/system} 

\subsection{Offline Lemma Synthesizer}  \label{sec:approach:offline_discovery}
% Goal of this subsection: explain how we bridge program semantics (i.e. intent) and tool VCs (i.e. from Frama-C) and how that produces the first batch of (offline) helper lemmas.

To generate helper lemmas that are aligned with the program semantics, we need to have a better understanding of the program semantics from the source code and specification, as well as how the semantics are reflected in the proof obligations.
The following subsections explain how we achieve them through an agentic program semantic analyzer and an obligation-aligned lemma synthesizer.

\subsubsection{Program Semantic Analyzer}
\label{sec:approach:offline_discovery:psa}

% why we need PSA
In traditional deductive program verification, human proof engineers usually manually write formal proof obligations based on program semantics and representations \cite{dodds2013mostly, nguyen2008enhancing}. 
Inspired by this tradition, we ask an LLM to generate a proof obligation by utilizing its capability both in code comprehension~\cite{nam2024using} and formal reasoning~\cite{lu2025palm}.
Such a proof obligation can be captured in an \emph{semantics-aware VC} that is expressed in terms of source-level concepts and is amenable to straightforward proof, which can be easily discharged by existing theorem proving techniques or by simply consulting an LLM.

% how PSA works in general
To this end, we design an agentic program semantic analyzer (PSA) that performs a comprehensive analysis of the code to be verified, with the prompt template shown in \autoref{fig:prompt-psa}. 
The prompt provides detailed information such as the source code, specifications (e.g., pre/post-conditions, loop invariants, and assertions), and other metadata to derive a semantics-aware VC that captures the high-level semantics of the program and the proof obligation.
To ease the practical concerns of input size and to focus the analysis on the most relevant parts of the code, we selectively examine portions of the code that are most pertinent to the proof obligations via slicing.
We instruct the LLM to document its understanding of the target property as a complete file in \rocq, including the statement \emph{and} its proof with all the supporting definitions.
This is possible because the semantics-aware VC is expressed in terms of source-level concepts and is amenable to straightforward proof. 
We apply a refinement process of up to five iterations to fix syntax errors or incorrect proofs until the output can be compiled with the \rocq compiler.
The semantics-aware VC is eventually given to the LLM as part of the prompt to produce potentially useful helper lemmas (see \autoref{sec:approach:discovery}) whose proofs are constructed separately. As such, we note that the semantics-aware VC only serves as \emph{guidance}, and its correctness does not affect the soundness of our approach. We ensure unproven statements are never imported into the proving context.
In the example from \autoref{fig:hex2bin}, the PSA would analyze the loop structure and the specifications to produce the semantics-aware VC as shown in Figure~\ref{fig:ghost_vc}) that captures the key invariant that \texttt{src} is monotonically increased by two in each iteration.
% The PSA design is justified for two reasons: (1) it can leverage the powerful reasoning capabilities of LLMs to perform deep semantic analysis of the code, and (2) it ensures the validity of the generated obligation with an accompanying proof, providing a solid foundation for the subsequent lemma synthesis process.

\input{asset/prompt-psa}

\subsubsection{Obligation-aligned Lemma Synthesis}
\label{sec:approach:discovery}

% why need obligation-aligned lemma synthesis
The PSA component generates a semantics-aware VC that is guaranteed to be valid as it is checked by the proof assistant.
However, there is no formal argument that the LLM-generated semantics-aware VC is faithful to the program semantics, and reasoning exclusively about this VC is inherently \emph{unsound}. 
As such, helper lemmas are synthesized to bridge the gap between the semantics-aware VC and the proof-targeted VC, so that the verifier ultimately discharges the VC produced by the VC generator.

To do so, we design an obligation-aligned lemma synthesis strategy that generates helper lemmas that are specifically tailored to bridge the representation gap between the two VCs, thus simplifying the proving of the proof-targeted VC while ensuring soundness.
Figure~\ref{fig:prompt-offline} illustrates the prompt template for this process. 
The input includes both the semantics-aware VC and the proof-targeted VC. In principle, these two lemmas should have encoded the same program and the same property, albeit produced by different means.
The LLM is instructed to compare them and output a set of helper lemmas that are aligned with both VCs and can be used to discharge the proof-targeted VC while remaining faithful to the program semantics captured by the semantics-aware VC.
Similar to the PSA component, here the LLM is also required to generate a complete \rocq file, \ie each helper lemma being proposed comes with a formal argument of correctness.
For example, given the input of the semantics-aware VC $\phi_A$ and the proof-targeted VC $\phi_C$ from \autoref{fig:hex2bin}, the agent synthesizes helper lemmas such as those shown in Figure~\ref{fig:helper_lemma}. 
To avoid generating an excessive number of trivial lemmas, we also export a proof plan that details how each helper lemma should be used to discharge the proof-targeted VC.

We note that the LLM may fail to produce a compilable \rocq file during offline synthesis for properties that involve sophisticated reasoning.
When that happens, we discard the lemmas causing compilation errors and recursively remove those dependent on them.
All the remaining well-formed helper lemmas and the proof plan are imported into the context of the proof agent to assist the online proving process.

\input{asset/prompt-offline}

\subsection{Online Lemma Adapter} \label{sec:approach:online_refinement}

% why need online lemma adapter
The lemmas synthesized in the offline synthesizer may not necessarily be sufficient to discharge the original verbose VCs.
As the synthesizer is not able to \emph{a priori} predict how the proof unfolds, the discovered lemmas may become inapplicable as the proof states evolve. 
The following subsections explain how we design an online lemma adapter that iteratively refines the helper lemmas and their supporting obligations to accommodate evolving proof states during the proving process, by either strengthening the lemma statement or revising the conflicting representation that the lemma depends on through feedback-guided lemma adaptation.

\subsubsection{Adaptive Lemma Maintainer} \label{sec:approach:alm}

% why we need ALM and how it works
During the proving process, certain important information (\eg retrieved helper lemmas) may be potentially useful for discharging the original, verbose VCs, but they are not directly applicable due to various reasons.
The adaptive lemma maintainer (ALM) is responsible for maintaining a corpus of helper lemmas and providing them to the agent during the proof process.
The helper lemmas maintained can be from various sources, including the initial lemmas synthesized from the previous section, the lemmas proved in the history proofs that these lemmas depend on (often implicitly through the program's control/data dependencies and specification structure), and the refined lemmas from the feedback-guided lemma adaptation in the next section.
In principle, other sources (e.g., program source code and specification) can also be provided to the agent as additional information for lemma refinement.
Here we particularly focus on the lemmas maintained in ALM as they are more directly related to the proof states and can be more easily reused and refined based on the evolving proof context.

% We also demonstrate that the program source code and specification can sometimes be too verbose and may not be directly useful for lemma refinement, as they may contain a lot of irrelevant information that can distract the agent from the key semantic insights needed for lemma refinement (see more details in \autoref{sec:evaluation:ablation}).

\input{asset/prompt-online}

\subsubsection{Feedback-guided Lemma Adaptation} \label{sec:approach:fla}

% why we need feedback-guided lemma adaptation and how it works
Inapplicable lemmas can occur due to type mismatch, \ie a helper lemma defined over one type (e.g., integer {\tt \small Z}) is inapplicable on terms of another type (e.g., {\tt \small addr}). Another common reason is naming conflict, where definitions in the helper lemmas are inconsistent with the ones in the proof state, \eg {\tt \small isint32} is defined with $\leq$ other than a strict inequality.
However, these inapplicable lemmas still provide valuable insights and can often become usable with minor adjustments. As such, we design a feedback-guided lemma adaptation approach that refines the helper lemmas and their supporting obligations to accommodate the evolving proof states.
Concretely, with access to the helper lemmas maintained in ALM and the current proof state, the agent proceeds with the following three strategies accordingly.

First, the best-case scenario is when the existing helper lemmas in ALM can be applied to the proof-targeted VC. In that case, since the helper lemmas are aligned with both the program semantics and the proof-targeted VC, the agent attempts to apply them directly.

Second, the application of these lemmas may fail due to the gap between the semantics-aware VC and the proof-targeted VC.
In this case, the agent refines the helper lemmas based on the feedback from the proof process, by either strengthening the lemma statement or revising the conflicting representation, and then attempts to apply the refined lemmas to discharge the proof-targeted VC.
This is the common case we observed in practice: the initial helper lemmas synthesized in the offline phase may not be directly applicable for normalization or on-the-fly induction, and may require refinement based on the evolving proof states to effectively discharge the proof-targeted VCs.
Concretely, Figure~\ref{fig:prompt-online} shows the prompt template used for the lemma adaptation step, which guides the agent to refine the helper lemmas and their supporting obligations based on the feedback from the proof assistant.

Third, if no lemmas in ALM can be applied to discharge the proof-targeted VC, the agent may propose new helper lemmas that are most relevant to the proof state, which can be synthesized from the agent's understanding of the proof progress, and then attempt to apply them to discharge the proof-targeted VC.
Through this iterative process of lemma refinement and application, the agent can effectively discharge complex VCs that are otherwise intractable, while maintaining the alignment with the program semantics captured by the semantics-aware VC.

\noindent\paragraph{Running Example.}
To recap the overall design of \toolname as shown in \autoref{fig:system}, we walk through the example from \autoref{fig:vc-cmp} again to illustrate how the offline lemma synthesizer and the online lemma adapter work together to discharge the proof-targeted VC.
In addition to feeding the source code and specification to the VC generator (\ding{192}$\to$\ding{193}), \toolname involves an offline lemma that digests these artifacts through the PSA component to derive the semantics-aware VC $\phi_A$, and then refers to $\phi_C$ to synthesize helper lemmas to bridge the representation gap. These lemmas are then added to the proof context to assist the proof agent (\ding{196}), and are collected by the adaptive lemma maintainer.
Then, during the proving process, the online lemma adapter iteratively refines the helper lemmas based on the feedback from the proof assistant so that lemmas become applicable in the evolved proof context (\ding{197}).
Collectively, the two stages of lemma discovery systematically encode semantic information from the program, and enable the proof agent to retrieve them on-the-fly as reusable artifacts.
%As a result, \toolname is sound (see \autoref{sec:discussions}) verification framework that effectively discharges complex VCs that are otherwise challenging (see \autoref{sec:evaluation:efficacy}). 
%We defer further discussions on how helper lemmas are proposed and used to \autoref{sec:evaluation:understanding_helper_lemmas}.

\subsection{Implementation}
We implement \toolname on top of an existing proof agent, \autorocq~\cite{autorocq}, in Python. The system is dependent on the \rocq proof assistant and the \framac ecosystem, specifically, the ACSL~\cite{DBLP:journals/fac/KirchnerKPSY15} and the WP plugin~\cite{wp}. 
The offline lemma synthesizer is implemented as a separate module that interacts with an LLM and checks its output in the \rocq proof assistant. The offline synthesizer simply imports generated helper lemmas and a proof plan into the context of the proof agent without interacting with it directly.
The online lemma adapter is implemented as a dynamic tool for the LLM~\cite{toolcall}. 
The proof agent interacts with it on demand by invoking the function and supplying the necessary arguments. Refined lemma statements are inserted as {\tt \small assert} embeddings in the proof context, and require separate subproofs to be constructed by the proof agent.
The integration of these components allows \toolname to effectively discover and apply helper lemmas that are crucial for proving complex VCs. %significantly improving the efficacy of automated program verification.
The underlying LLM for the proof agent and all lemma discovery steps is configurable and is set to gpt-5.2-2025-12-11 with temperature 0 for better reproducibility.

%% file: figures/system.tex
\begin{figure}[t]
	%\centering
	\includegraphics[width=1\linewidth]{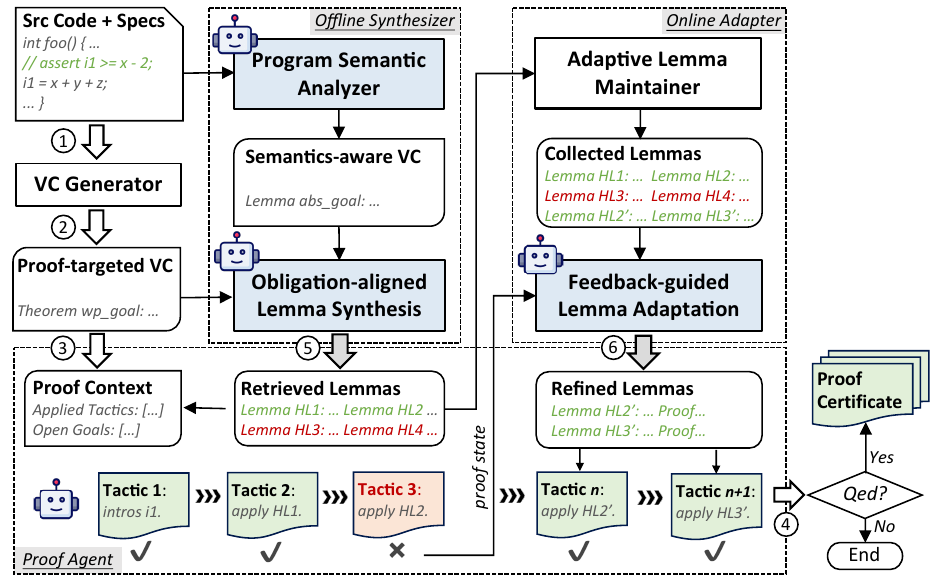}
    %\vspace{-1em}
	\caption{System overview of \toolname. In addition to a typical workflow that directly reasons about and proves the VC produced by a VC Generator (\ding{192}$\to$\ding{193}$\to$\ding{194}$\to$\ding{195}), \toolname adopts (1)~an offline lemma synthesizer that extracts helper lemmas directly from the source code and specifications \ding{196}, and (2)~an online lemma adapter that refines helper lemmas based on the proof state and existing lemma contexts \ding{197}.}
	\label{fig:system}
\end{figure}

%% file: asset/prompt-psa.tex
\begin{figure}
\begin{promptresponsebox}[]
{\bf \textit{Prompt:}}
Your task is to analyze the annotated source code with ACSL annotations and write a complete \rocq file that intuitively proves what the annotation states. Focus only on the highlighted property to prove... Available context: 
\begin{itemize}[nosep]
    \item Property name: {\tt \small hex2bin\_loop\_invariant\_2}
    \item Location: function {\tt \small hex2bin} at line 10 of file {\tt \small hex2bin.c}
    \item Annotated source code: <\autoref{fig:hex2bin}>
\end{itemize}

{\bf \textit{Response:}} Here is the \rocq file: <Figure~\ref{fig:ghost_vc}>
\end{promptresponsebox}
\vspace{-1em}
\caption{Prompt for program semantic analyzer.}
%\vspace{-1em}
\label{fig:prompt-psa}
\end{figure}

%% file: asset/prompt-offline.tex
\begin{figure}[t]
%\vspace{-0.5em}
\begin{promptresponsebox}[]
{\bf \textit{Prompt:}}
Analyze an \emph{equivalent} goal $\phi_C$ that has been directly discharged from \framac using the same code and annotations: <Figure~\ref{fig:original_vc}>. Based on the previously proved lemma $\phi_A$, propose \emph{strong enough helper lemmas} to help prove $\phi_C$. Note that you \emph{do not} need to prove $\phi_C$ itself... Important guidelines:

\begin{enumerate}
    \item For each helper lemma proposed, provide its proof.
    \item Helper lemmas can be very specific to the goal (e.g., you may use specific constants).
    \item Describe a step-by-step plan detailing where each helper lemma can be applied.
\end{enumerate}

{\bf \textit{Response:}} Here are the helper lemmas: <Figure~\ref{fig:helper_lemma}>
% \begin{lstlisting}[language=Coq,frame=none,xleftmargin=0.12\linewidth]
%   Lemma HL1: ... Proof...Qed.
%   Lemma HL2: ... Proof...Qed.
%   Lemma HL3: ... Proof...Qed.
%   Lemma HL4: ... Proof...Qed.
% \end{lstlisting}

\end{promptresponsebox}
\vspace{-1em}
\caption{Prompt for obligation-aligned lemma synthesis.}
\label{fig:prompt-offline}
%\vspace{-1em}
\end{figure}

%% file: asset/prompt-online.tex
\begin{figure}
%\vspace{-0.5em}
\begin{promptresponsebox}[]
{\bf \textit{Prompt:}}
Here is the current proof state:
\begin{enumerate}
    \item Applied tactics: \lstinline[language=Coq]{Proof. intros i1. apply HL1.} 
    \item Open goal: \lstinline[language=Coq]{forall i i1 x y: int, ...}
    \item Error feedback: Tactic \lstinline[language=Coq]{apply HL2.} failed because term  \texttt{HL2} has type {\tt x<i} while it is expected to have type {\tt x<i1}.
\end{enumerate}

Here is a list of helper lemmas. If a critical lemma exists but is not applicable, propose a refined version with corrected types:
\begin{lstlisting}[language=Coq,frame=none,xleftmargin=0.12\linewidth]
  Lemma HL1: ... Proof...Qed. (* imported *)
  Lemma HL2: ... Proof...Qed. (* imported *)
  Lemma HL3: ... Proof...Qed. (* conflict *)
  Lemma HL4: ... Proof...Qed. (* conflict *)
\end{lstlisting}
{\bf \textit{Response:}} Here is the refined lemma: \lstinline[language=Coq]{Lemma HL2': ...}
\end{promptresponsebox}
\vspace{-1em}
\caption{Prompt for feedback-guided lemma adaptation.}
\label{fig:prompt-online}
\end{figure}

%% file: sections/evaluation.tex
\section{Evaluation} \label{sec:eva}

\subsection{Evaluation Setup} \label{sec:eval:setup}

To evaluate the efficacy of \toolname, we design a comprehensive evaluation that answers the following research questions (RQs):

\begin{enumerate}
    \item[{\bf [RQ1]}] Is \toolname effective in proving verification conditions from real-world C programs? 
    \item[{\bf [RQ2]}] How does each component of \toolname contribute to its overall efficacy?
    \item[{\bf [RQ3]}] To what extent are the helper lemmas useful? How are they used in successful proofs?
\end{enumerate}

\noindent\paragraph{Benchmarks.} 
We evaluate \toolname on a total of 941 verification conditions (VCs) extracted from real-world C programs, which are collected from two existing sources:
\begin{itemize}[leftmargin=1em]
    \item 641 VCs extracted from the \textbf{SV-COMP} benchmarks~\cite{svcomp}, which have been used in the evaluation of \cite{autorocq}. These VCs are extracted from 131 C programs, with an average size of 43 lines of code. The largest base program is a BusyBox utility spanning 428 lines.
    \item 300 VCs from the test partition of \textbf{NTP4VC}~\cite{xu2026neural}, the largest benchmark to date in real-world verification conditions. 
    Specifically, these theorems are collected from eight critical C projects, including the Linux kernel (scheduler~\cite{lawall2024should}, memory management \cite{efremov2018deductive}, and string utilities \cite{carvalho2014formal}), Contiki OS~\cite{blanchard2018ghosts}, C++ standard library \cite{Burghardt_Gerlach_Lapawczyk_2015}, X.509 parser~\cite{ebalard2019journey}, the  UAV autopilot~\cite{pollien2021verifying}, and more. 
    These programs are considerably larger than the SV-COMP benchmarks, with several exceeding 1,000 lines of code. The largest among them, X.509 parser, comprises 5,044 LoC.
\end{itemize}
Proving VCs from these benchmarks involves reasoning about various language constructs, such as pointers, floating-point arithmetic, and custom data structures.
In particular, theorems from NTP4VC tend to be much more complex than those from SV-COMP due to the larger program sizes, posing significant challenges to existing approaches~\cite{xu2026neural}.
Regarding the type of properties being proved, the 941 real-world VCs can be broadly grouped into four categories. A large portion of them (391, 41.6\%) are related to loops, including loop invariants/variants and loop assignments; 237 (25.2\%) VCs specify RTE-freeness properties such as non-overflow and valid memory accesses; 163 (17.3\%) VCs encode functional correctness through assertions; finally, 150 (15.9\%) VCs are extracted from functional contracts, including pre/post-conditions and behavioral partitioning (e.g., {\tt \small disjoint} and {\tt \small complete}).
Collectively, these benchmarks form an ideal testbed for evaluating VC proving tools.

\smallskip
\textit{Comparative Approaches.} We compare \toolname with the following approaches that are applicable for proving program VCs:
\begin{itemize}[leftmargin=1em]
    \item {\bf \autorocq}~\cite{autorocq}, an LLM agent that collaborates with the \rocq ITP and retrieves additional context on demand to conduct proof search. It is the base proof agent upon which \toolname is built.
    \item {\bf \copra}~\cite{copra24}, an LLM agent that generates proofs through in-context learning, searching, and knowledge retrieval.
    \item {\bf \hammer}~\cite{coqhammer}, a popular tool that uses machine-learned heuristics and SMT solving to automate theorem proving.
\end{itemize}
We include \autorocq and \copra as they are state-of-the-art LLM-based theorem provers, as evidenced by prior work~\cite{autorocq}. 
We include \hammer as the non-agentic baseline that has been reported to outperform naive LLM-based approaches~\cite{xu2026neural}.
We additionally include three ablative variants of \toolname in \autoref{sec:evaluation:ablation} to evaluate the contribution of individual components.
Unfortunately, existing tools in automated lemma discovery~\cite{sivaraman2022lfind, brendel2025synthesizing, mccasland2017mathsaid} target mathematical theorems only, and cannot be applied in our setting. We delay the detailed discussion and qualitative comparison to \autoref{sec:related}.
In our evaluation, we use the same backend LLM gpt-5.2-2025-12-11 for all tools, with temperature set to 0. \toolname, \autorocq, and \copra are run for a budget of 10 minutes or 100 steps, following the setup in \cite{autorocq}. \hammer, which does not involve an LLM, is run with the default setting and the same 10-minute timeout.

\subsection{RQ1: Efficacy in VC Proving} \label{sec:evaluation:efficacy}
For this research question, we investigate \toolname's effectiveness in proving real-world verification conditions. We report the number of proved VCs from each comparison tool
in \autoref{tab:results-all}. 

For the SV-COMP benchmark, \toolname is able to prove 298 out of 641 VCs, outperforming the best-performing baseline, \autorocq, by 20.6\%. 
\copra is able to prove 209 VCs. \hammer only proves 85 VCs within the 10-min timeout, lagging behind three agent-based proving techniques by a large margin.
In contrast, the success rates across the NTP4VC benchmark drop across all tools due to increased complexity in the VC statements.
\toolname is able to prove 66 out of 300 VCs, outperforming \autorocq (40) substantially by 60\%. \autorocq is closely followed by \hammer, which manages to prove 38 VCs. 
Across all evaluated VCs, \toolname improves over comparison tools by 26.8\%--195.9\%, demonstrating its remarkable efficacy in proof generation relative to existing approaches.

In addition, we present \autoref{fig:succ_venn}, a Venn diagram that visualizes the relation among the set of VCs proved by each tool. The results are aggregated across both benchmarks. 
The diagram shows that \toolname proves the largest number of VCs uniquely (78), significantly outperforming comparison tools.
\toolname is followed by \hammer, which obtains 26 unique successes despite proving the least number of VCs overall, highlighting the ability
of SMT-based approaches to effectively complement agentic approaches.

\input{figures/results}

\input{figures/results-breakdown}

To better understand the efficacy of \toolname, we categorize the successfully proved VCs along two dimensions and report the resulting breakdown for each tool in \autoref{fig:succ_breakdown}.
Figure~\ref{fig:succ_by_complexity} correlates the number of proved VCs with their complexity. Here, the complexity of a VC is measured by the number of terms contained in the theorem statement, and a larger count (counts increase from left to right on the $x$-axis) indicates higher complexity \cite{autorocq}. 
On the other hand, Figure~\ref{fig:succ_by_type} groups proved VCs by the type of property they denote, as discussed in \autoref{sec:eval:setup}.
In both plots, \toolname consistently leads across \emph{all} buckets.
These results highlight the effectiveness
of \toolname in comparison with other tools in proving VCs from real-world programs. The results also indicate that helper lemma discovery is widely applicable across the full range of VC complexities and property types.

\begin{resultbox}
{\bf Answer to RQ1}: In comparison with the other tools in the study, \toolname is highly effective in proving VCs from real-world programs. 
%It proves the most VCs uniquely and consistently outperform comparison tools across the full range of VC complexities and VC types.
It uniquely proves the largest number of VCs and consistently outperforms comparison tools across the full range of VC complexities and types.
\end{resultbox}

\subsection{RQ2: Ablation Studies} \label{sec:evaluation:ablation}

To answer RQ2, we conduct ablation studies to understand how much each \toolname component contributes to the overall effectiveness by comparing \toolname with the following variants:

\begin{itemize}[leftmargin=1em]
    \item \textbf{\varNoOffline}, a variant with only the online adapter, i.e., the offline synthesizer is disabled;
    \item \textbf{\varNoOnline}, a variant with only the offline synthesizer, i.e., the online adapter is disabled;
    \item \textbf{\varNaiveOffline}, a variant that removes the program semantic analyzer (PSA) component, i.e., offline helper lemmas are generated directly from the proof-targeted VC without semantic knowledge about the annotated source code.
    % \item \textbf{\varProgOnline}, a variant without the offline synthesizer, but the online adapter has access to the annotated source code as program information
    % \item \textbf{\varNoALM}, a variant with the default online lemma proposal, i.e., with no information from the ALM component performed.  
    % \item \textbf{\varALML}, a variant with default online + existing lemmas (offline lemmas + refined lemmas)
    % \item \textbf{\varALMP}, a variant with the default online + helper\_lemma (check) of the source code 
    % \item \textbf{\varALMLP},  a variant with a variant with the default online + helper\_lemma (check) of the source code  + existing lemmas (offline+refined)
\end{itemize}
Each of the first two variants, $\neg$Ofl and $\neg$Onl, individually disables one stage of the lemma discovery process, namely \ding{196} and \ding{197} in \autoref{fig:system}.
We include the third ablative variant, $\neg$PSA, to evaluate the effectiveness of program-guided offline lemma discovery.

We evaluate each variant on the entire benchmark set of 941 VCs. The last section of 
\autoref{tab:results-all} presents results from these variants. These results show that all \toolname components make meaningful contributions to its overall effectiveness. 
In particular, the comparison with variant \varNaiveOffline highlights the importance of program comprehension in generating semantics-aligned helper lemmas that are conducive to online proving.

\begin{resultbox}
{\bf Answer to RQ2}: The offline lemma synthesizer, the online lemma adapter, and the program semantic analyzer all contribute to the overall effectiveness of \toolname.
% The offline synthesizer provides a strong foundation of helper lemmas that capture the program semantics, while the online adapter further refines these lemmas and discovers new ones on-the-fly.
\end{resultbox}

\subsection{RQ3: Understanding Helper Lemmas} \label{sec:evaluation:understanding_helper_lemmas}

We next delve into the details of the helper lemmas to better understand the lemmas that are proposed and how they are used in successful proofs.
To this end, we first report aggregated statistics for \toolname's lemma discovery phases along four dimensions, and then present two case studies on how they are used in the corresponding proofs.

\subsubsection{Quality}
We first study \emph{how good} the discovered helper lemmas are.
Unfortunately, a direct, similarity-based measure of their quality~\cite{sivaraman2022lfind, brendel2025synthesizing} is infeasible in this case due to the lack of human-written ground truths.
As such, we resort to an indirect measure by counting how many successful proofs make use of the proposed lemmas (e.g., via {\tt \small apply} or {\tt \small rewrite}).
We report the counts in \autoref{fig:hl_venn} for all 364 VCs proved by \toolname and 78 uniquely proved VCs.
Among the 78 unique successes, only 9 (11.5\%) of the proofs do not require an offline or online lemma. 32 (41.0\%) require offline, but not online lemmas, 17 (21.8\%) require online, but not offline lemmas, and 20 (25.7\%) require both kinds of lemmas. These numbers highlight the usefulness of these lemmas.
The percentages are lower when considering all successes, as many simpler theorems can be directly proved without relying on helper lemmas. 
Overall, offline lemmas synthesized tend to be used more often, indicating the importance of program awareness in lemma discovery.

\input{figures/hl_usage}

\input{figures/hl_stats}

\subsubsection{Cardinality}
We next report statistics on \emph{how many} helper lemmas are discovered by \toolname and subsequently used in successfully proved VCs (max.\ 212=57+95+60).
We present the results in \autoref{fig:hl_hist}, where the $x$-axis lists the VCs ordered by the number of lemmas discovered (\ie $y$-axis).
A few patterns are visible from the diagram. 
First, in general, more helper lemmas are discovered in the offline stage (blue) compared to the online phase (orange). 
Second, discharging the proof-targeted VC typically only requires a few key helper lemmas, as only a small fraction of lemmas discovered offline are being directly used in proofs.
Third, we observe that the lemmas discovered online, although fewer than offline lemmas, are often immediately useful to the proof.
This is expected as the offline discovery phase is unaware of proof states, whereas the online adaptation is driven by real-time feedback from \rocq. 

\subsubsection{Utility Taxonomy}
We report \emph{what purpose} the helper lemmas serve in \autoref{tab:hl_cat} by categorizing them based on keyword matching. We report the categorization results for (1) the list of all helper lemmas discovered, and (2) the ones being used in successful proofs.
The most common category for both cases is memory-related helper lemmas that contain keywords such as {\small \tt addr} and {\small \tt ptr}, with Figure~\ref{fig:helper_lemma} being concrete examples. 
28.1\% of all discovered lemmas and 30.2\% of used lemmas fall into this category. 
This is followed by simplification lemmas that rewrite, reduce, or split a large theorem into smaller or simpler (sub-)goals that are easier to reason about.
Other popular categories include typing bridges (e.g., mapping a {\tt \small uint32} variable to its range as integer {\tt \small Z}) and arithmetic lemmas (e.g., expressing left-shifting as multiplication), each accounting for over 10\% of the lemmas.

\subsubsection{Proof Strategy Taxonomy}
We further inspect all VCs that are uniquely proved by \toolname with the help of helper lemmas (max.\ 69=20+32+17). We manually categorize how these successful proofs make use of helper lemmas based on the proof strategies they enable.
We identify four primary strategies: 
\begin{itemize}[leftmargin=1em]
\item \emph{BRIDGE}: connecting representations across semantic levels,
\item \emph{REDUCE}: search-space reduction,
\item \emph{REWRITE}: normalization or term rewriting, and
\item \emph{STRENGTHEN}: deriving stronger usable facts.
\end{itemize}
Among the 69 proofs that utilize helper lemmas, BRIDGE is the most prevalent, appearing in 61 proofs (88.4\%), underscoring the critical role of helper lemmas in connecting tool-specific representations to program-level semantics.
REDUCE follows closely at 40 proofs (58.0\%), while REWRITE and STRENGTHEN appear in 34 (49.3\%) and 33 (47.8\%) proofs, respectively.
Other strategies such as case-splitting, contradiction-exposing, and modularization are used less frequently (under 6\% each).
Notably, most proofs employ multiple strategies---averaging 2.5 per proof---demonstrating that effective VC proving requires helper lemmas serving complementary roles.

\input{figures/hl_category}

\subsubsection{Case Studies}
We present representative case studies that illustrate how helper lemmas facilitate VC proving. 
The theorem statements (\ie proof-targeted VCs) have been simplified for clarity.
%Each case follows a consistent narrative:
%We describe the origin of the proof goal, the property it encodes, the helper lemmas involved, and why these lemmas enable the proof to succeed.

\smallskip
\noindent
\textit{Case I: Refined helper lemma from failed proofs (Listing 1).}
% should_we_balance_assert_rte_mem_access_12_goal23
This VC from the Linux kernel scheduler requires proving {\tt \small valid\_rd t a\_5 1}, where {\tt \small a\_5 = shift a\_4 0}. Although {\tt \small valid\_rd t a\_4 1} is in the hypotheses, \rocq cannot automatically conclude validity for {\tt \small a\_5}. The online adapter synthesizes {\tt \small HL\_valid\_rd\_shift0} when it fails to apply {\tt \small HL\_valid\_rd\_rewrite}, stating that shifting by zero preserves memory validity, bridging the gap and completing the proof.

\smallskip
\input{figures/case-online-refined}

\smallskip
\noindent
\textit{Case II: Combining offline and online helper lemmas (Listing 2).}
%rocq_memb_alloc_Why3_ide_VCmemb_alloc_assert_rte_signed_overflow_2_goal11.v
This VC from a Contiki OS memory allocator requires proving {\tt \small -2147483648 <= i * x\_1} (no signed underflow), given {\tt \small 0 <= i} and {\tt \small is\_uint16 x\_1}. 
\toolname employs two lemmas: {\tt \small HL\_uint16\_nonneg} (from offline), extracting the fact {\tt \small 0 <= x\_1} from the type predicate, and {\tt \small HL\_mul\_nonneg\_of\_nonneg} (from online), concluding a non-negative product from non-negative factors. This illustrates offline-online synergy: domain-specific type knowledge combined with arithmetic reasoning.

\smallskip
\input{figures/case-combined}

\begin{resultbox}
{\bf Answer to RQ3}: The helper lemmas discovered by \toolname are highly useful in VC proving. They bridge representation gaps and provide key intermediate results, enabling reasoning steps that are otherwise inaccessible to the agent.
\end{resultbox}

%% file: figures/results.tex
\begin{table}[t]
    \centering
    \small
    \caption{Results on proving verification conditions by different approaches and ablative variants: the number of successfully proved VCs among SV-COMP (max. 641) and NTP4VC (max. 300) benchmarks, as well as the relative improvement ({Improv.}) of \toolname.} \label{tab:results-all}
    %\vspace{-1em}
    \setlength{\tabcolsep}{6pt}
    \begin{tabular}{l|c|c|cc}
        \toprule
        % \multirow{2}{*}{\textbf{Tools}} &
        % \multicolumn{2}{c|}{\textbf{SV-COMP}} &
        % \multicolumn{2}{c|}{\textbf{NTP4VC}} &
        % \multicolumn{2}{c}{\textbf{ALL}} \\
        % \cmidrule(lr){2-3}\cmidrule(lr){4-5}\cmidrule(lr){6-7}
        % & \textit{Num} \textit{Num} & \textit{Num} & \textit{Imp} \\
        \textbf{Tools} & \textbf{SV-COMP} & \textbf{NTP4VC} & \textbf{Total} & \textbf{Improv.} \\
        \midrule
        \toolname     & 298 & 66 & 364 & -- \\
        \midrule
        \autorocq \cite{autorocq}     & 247 & 40 & 287 & +26.8\% \\
        \copra \cite{copra24}         & 209 & 31 & 240 & +51.7\%  \\
      % \rango \cite{rango-icse25}    & 139 & 0  & 139 &  \\
        \hammer \cite{coqhammer}      & 85  & 38 & 123 & +195.9\% \\
        \midrule
        \varNoOffline           & 273 & 55  & 328 & +11.0\% \\
        \varNoOnline            & 246 & 56  & 302 & +20.5\% \\
        \varNaiveOffline        & 253 & 48  & 301 & +20.9\% \\
      % \varProgOnline  &  &  &  &  \\
        \bottomrule
    \end{tabular}
\end{table}

% \hlt{I would suggest using back the previous version with individual improvement for each benchmark, which makes the comparison comprehensive: we can discuss the improvements from sv-comp and ntp4vc separately. If the space is an issue, can use a double column to present, as this is an important comparison}. 

%% file: figures/results-breakdown.tex
%\noindent
\begin{figure}[t]
\centering
%\vspace{-1em}
\includegraphics[width=0.97\linewidth]{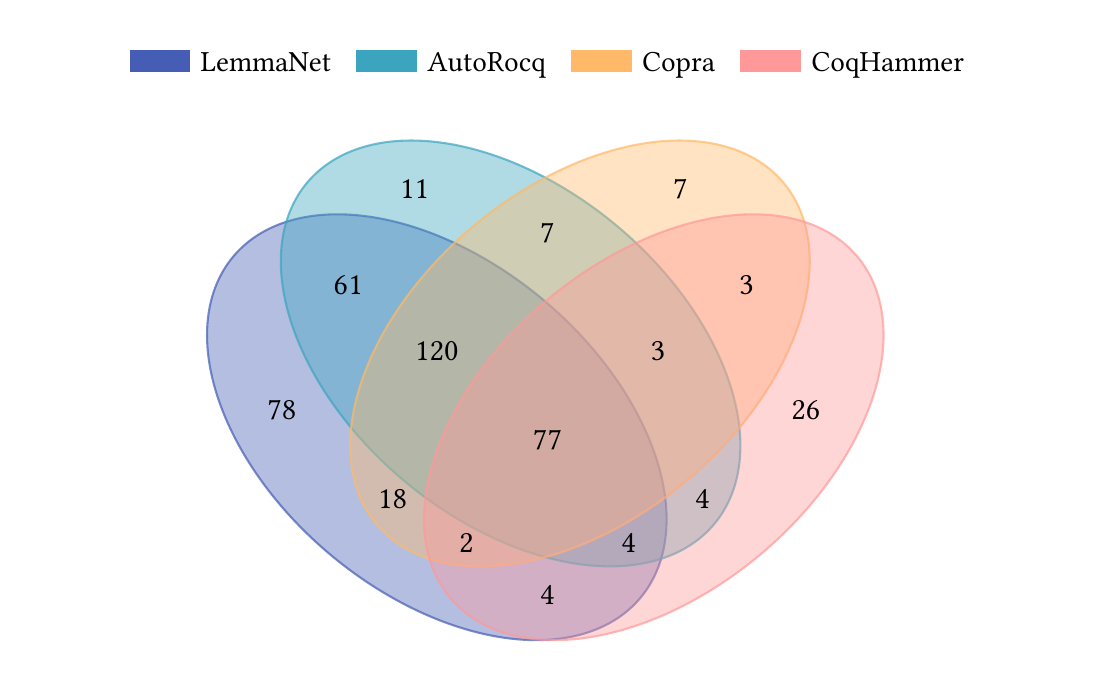}
%\vspace{-1em}
\captionof{figure}{Venn diagram of the number of lemmas proved by each tool. \toolname uniquely proves the most VCs.}
\label{fig:succ_venn}
%\vspace{-1em}
\end{figure}

% \ourSol is able to prove more lemmas, and has the most number of uniquely proved lemmas.

\begin{figure}[t]
%\vspace{-2em}
%\centering
\subfigure[Breakdown by term complexity.]{	
\begin{minipage}[t]{0.475\linewidth}
\centering
\includegraphics[width=1\linewidth]{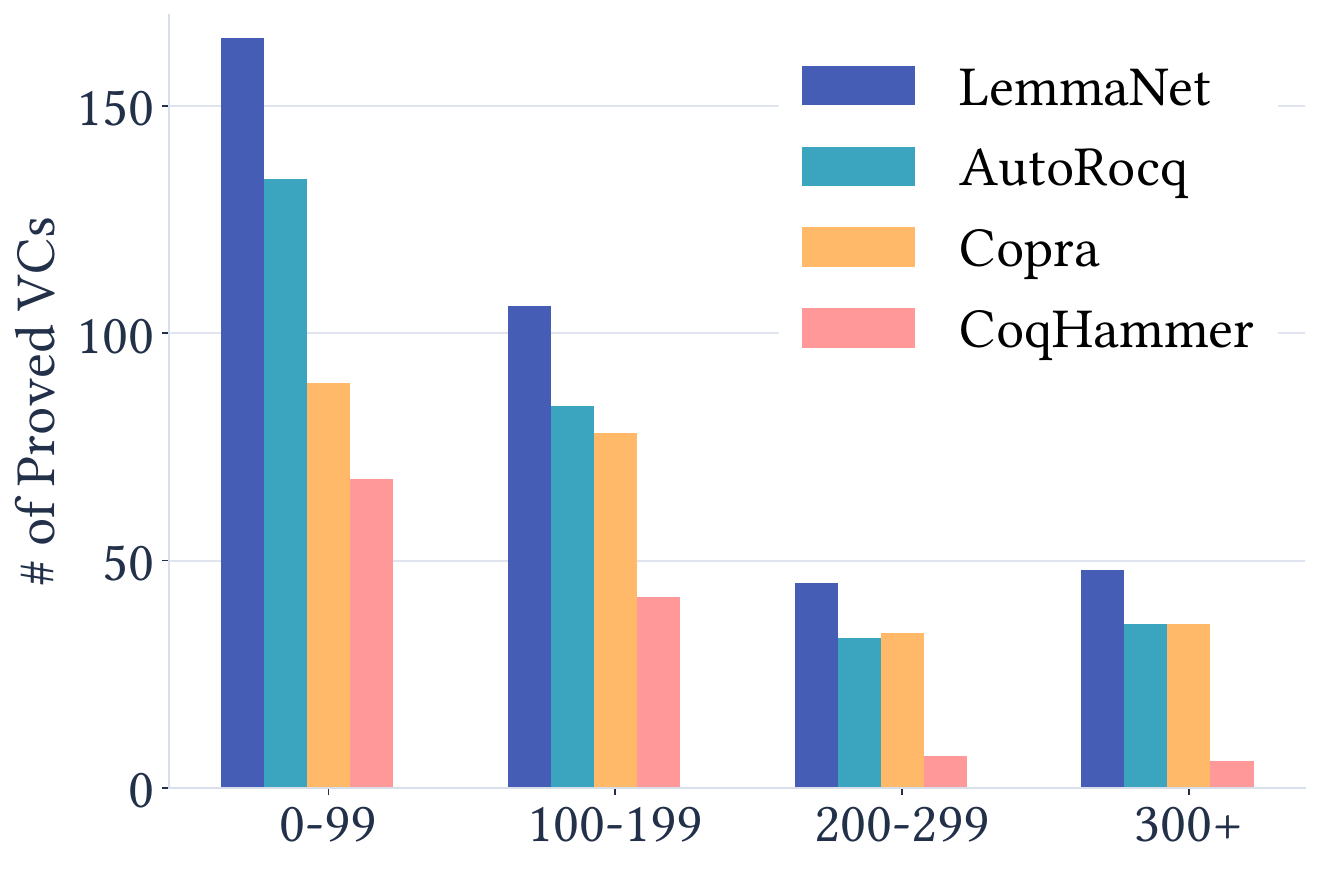}
\label{fig:succ_by_complexity}
\vspace{-1em}
\end{minipage}	
}
\hfill
\subfigure[Breakdown by property type.]{	
\begin{minipage}[t]{0.475\linewidth}
\centering
\includegraphics[width=1\linewidth]{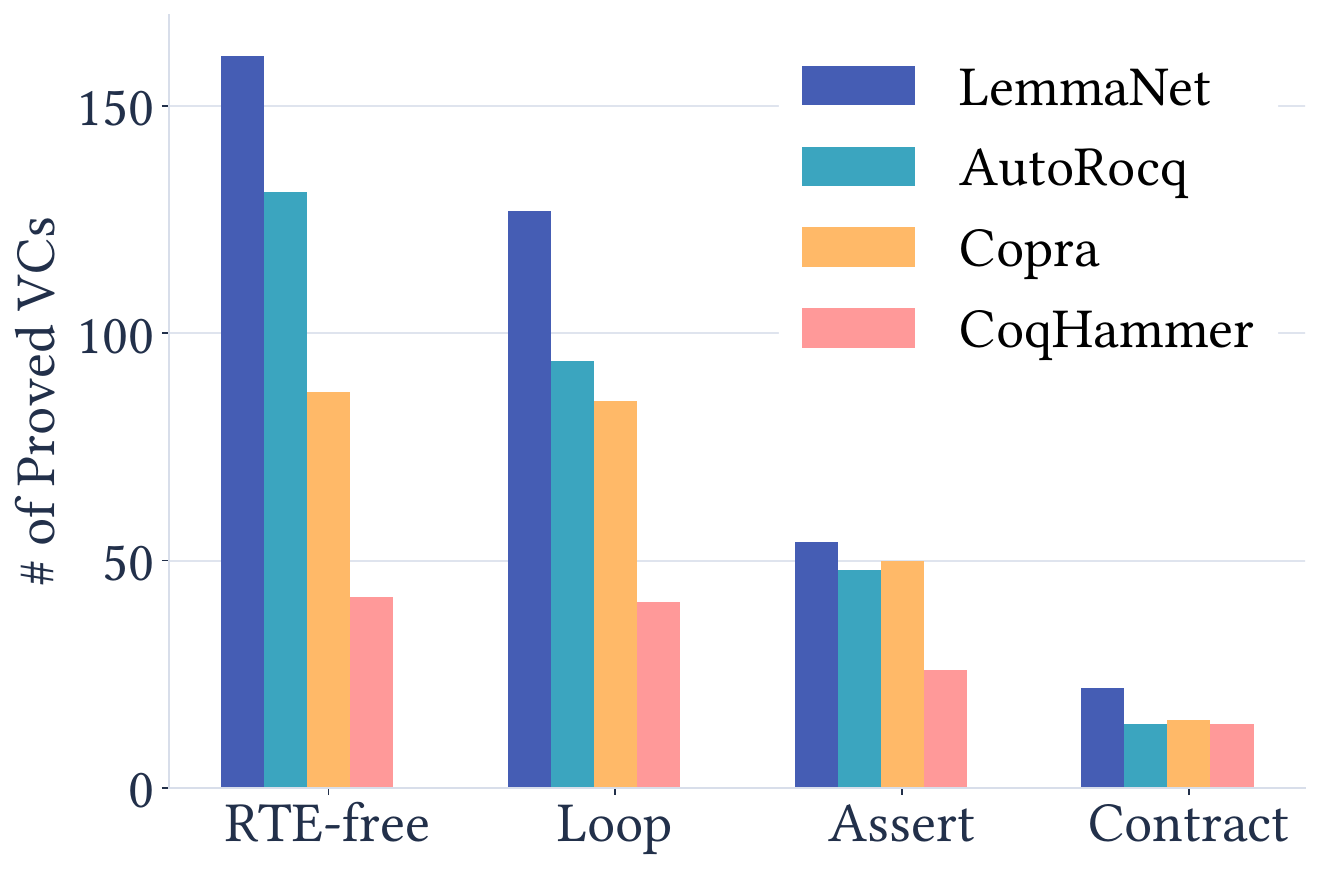}
\label{fig:succ_by_type}
\vspace{-1em}
\end{minipage}	
}
%\vspace{-1em}
\caption{The number of VCs proved by different tools: a breakdown by complexity (\# of terms) and property type.}

\label{fig:succ_breakdown}
%\vspace{-1em}
\end{figure}

%% file: figures/hl_usage.tex
% \begin{figure}[t]
% 	\centering
% 	\includegraphics[width=0.9\linewidth]{asset/hl-venn.pdf}
%     \vspace{-1em}
% 	\caption{Quality of helper lemmas: \# of successful proofs that make use of online/offline lemmas, reported for all (left) and unique VCs (right) proved by \toolname. 89.7\% of unique successes make use of \emph{either} type of lemmas.}
% 	\label{fig:hl_venn}
% \end{figure}

\begin{table}[t]
\centering
\caption{Quality of helper lemmas: \# of successful proofs that make use of both, either, or none of online/offline lemmas, reported for all and unique VCs proved by \toolname. }
\label{fig:hl_venn}
\begin{tabular}{c|ccc|c|c}
\toprule
 & \textbf{Both} & \textbf{Offline} & \textbf{Online} & \textbf{None} & \textbf{Total} \\
\midrule
\multirow{2}{*}{{\bf All}}
& 57 & 95 & 60 & 152 & 364 \\
& 15.7\% & 26.1\% & 16.5\% & 41.7\% & 100\% \\
\hline
\multirow{2}{*}{{\bf Unique}} 
& 20 & 32 & 17 & 9 & 78 \\
& 25.7\% & 41.0\% & 21.8\% & 11.5\% & 100\% \\
\bottomrule
\end{tabular}
\end{table}

%% file: figures/hl_stats.tex
\begin{figure}[t]
	\centering
    %\vspace{-1em}
  % \includegraphics[width=0.8\linewidth]{asset/hl-hist.pdf}
    \includegraphics[width=\linewidth]{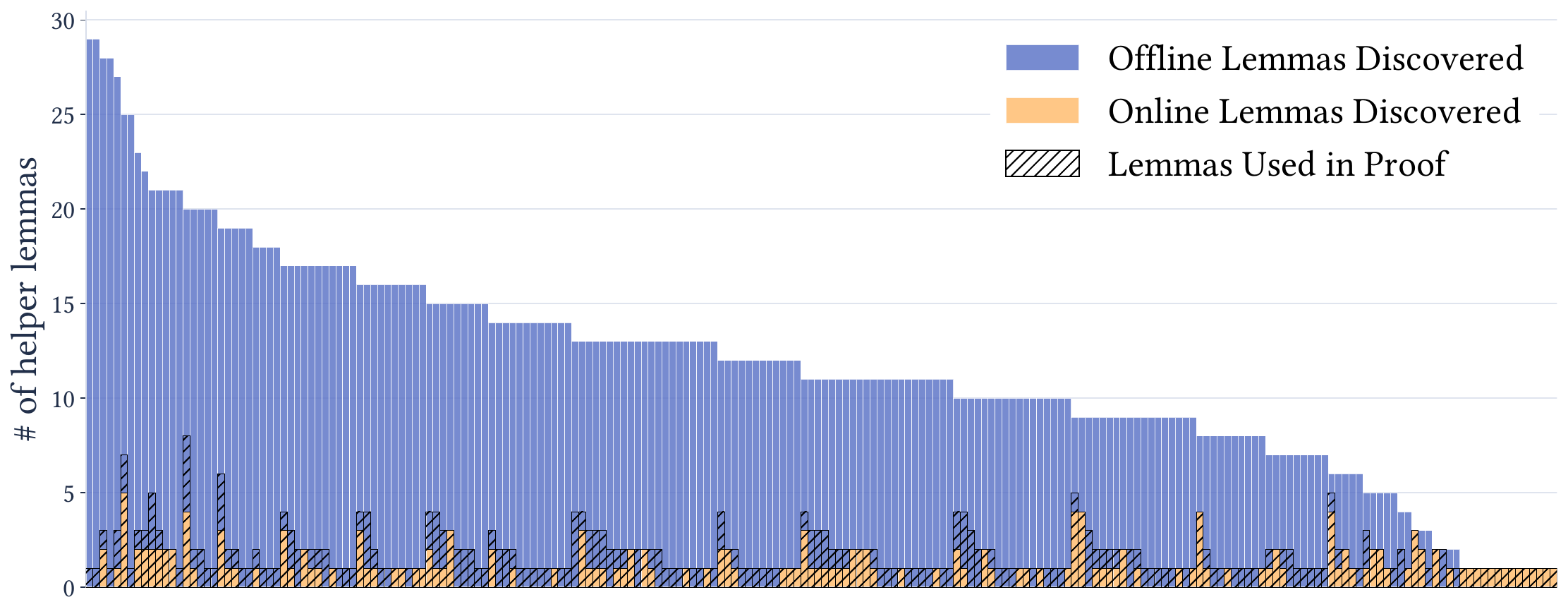}
    %\vspace{-1em}
	\caption{Cardinality of helper lemmas: \# of offline/online lemmas discovered and used for successfully proved VCs.}
	\label{fig:hl_hist}
    %\vspace{-1em}
\end{figure}

%% file: figures/hl_category.tex
\begin{table}[t]
    \centering
    %\small
    \caption{Utility taxonomy of helper lemmas: a breakdown by category. Results are reported for both discovered (\% Disc.) and used (\% Used) lemmas. Categories are determined through keyword matching on lemma names. %Memory-related helper lemmas are the most common.
    } 
    \label{tab:hl_cat}
    \setlength{\tabcolsep}{6pt}
    \resizebox{0.95\linewidth}{!}{
    \begin{tabular}{l|c|c|l}
        \toprule
        \textbf{Category} & \textbf{\% Disc.} & \textbf{\% Used} & \textbf{Example Keywords} \\
        \midrule
        Memory & 28.1\% & 30.2\% & {\small \tt addr, base, ptr, mem} \\
        Simplification & 25.6\% & 22.0\% & {\small \tt simpl, rewrite, split} \\
        Typing & 19.3\% & 18.8\% & {\small \tt uint32, float, INT\_MAX} \\
        Arithmetic & 13.5\% & 17.2\% & {\small \tt mul, div, mod, lxor} \\
        Data Structure & 7.8\% & 9.0\% & {\small \tt array, list, heap, map} \\
        String & 3.5\% & 2.4\% & {\small \tt str, char, tolower} \\
        Others & 2.2\% & 0.4\% & -- \\
        \bottomrule
    \end{tabular}
    }
\end{table}

%% file: figures/case-online-refined.tex
%\begin{figure}[t]
\begin{minipage}[c]{0.90\linewidth}
\begin{lstlisting}[language=Coq,xleftmargin=0pt,xrightmargin=0pt,linewidth=0.98\linewidth,
caption={Refined helper lemma from failed proofs.}]
Lemma HL_valid_rd_rewrite :
  forall (t : Z -> Z) (a b : addr) (n : Z), 
    valid_rd t a n -> a = b -> valid_rd t b n.
Proof. intros t a b n H Hab. eapply valid_rd_ext_addr; eauto. Qed.

Theorem wp_goal_reduced
  (a_1 : addr) (t_4 : addr -> addr) (t : Z -> Z) :
  let a_3 : addr := shift a_1 4%Z in let a_4 : addr := t_4 a_3 in
  let a_5 : addr := shift a_4 0%Z in valid_rd t a_4 1%Z ->
  valid_rd t a_5 1%Z.
Proof.
  intros; cbn.
  assert (HL_valid_rd_shift0: forall (t:Z->Z) (p:addr) (n:Z), 
      valid_rd t p n -> valid_rd t (shift p 0%Z) n). {
  intros t0 p n Hrd Hpos; destruct (Hrd Hpos) as [Hb [Hoff Hle]];
  repeat split; try assumption; simpl; lia. }
  exact (HL_valid_rd_shift0 t a_4 1 H22).
Qed.
\end{lstlisting}
\label{fig:case-study-2}
\end{minipage}
%\end{figure}

%% file: figures/case-combined.tex
%\begin{figure}[t]

\begin{lstlisting}[language=Coq,xleftmargin=0pt,xrightmargin=0pt,linewidth=0.98\linewidth,
caption={Combining offline and online helper lemmas.}]
Theorem wp_goal_reduced (a : addr) (i : Z) (t_1 : addr -> Z) :
  let x_1 : Z := t_1 (shift a 0%Z) in 0%Z <= i ->
  is_uint16 x_1 -> -2147483648%Z <= i * x_1.
Proof.
  intros; cbv zeta.
  assert (HL_Hx1_nonneg : 0 <= x_1) by (apply HL_uint16_nonneg; exact H9).
  assert (HL_Hmul_nonneg : 0<=i*x_1) by (apply HL_mul_nonneg_of_nonneg; 
      [exact H0 | exact HL_Hx1_nonneg]). lia.
Qed.
\end{lstlisting}
\label{fig:case-combined}
%\end{figure}

%% file: sections/discussion.tex
\section{Discussion} \label{sec:discussions}

\textit{Experimental Costs.}
We report the LLM API costs for running \toolname. For offline synthesis, the mean cost per VC is \$0.12 (SV-COMP) and \$0.17 (NTP4VC), with medians of \$0.10 and \$0.14, respectively. For the proof agent phase (which includes online adaptation), the mean cost is \$0.36 (SV-COMP) and \$1.03 (NTP4VC), with medians of \$0.07 and \$0.37. The higher costs for NTP4VC reflect its greater VC complexity. Overall, these costs are comparable to \autorocq~\cite{autorocq} and \copra~\cite{copra24}, and are modest given the significant improvements in proving capability.
The average proving time is 165.2 seconds per VC in our evaluation, making \toolname a practical tool for real-world verification tasks.

\smallskip
\textit{Soundness.} 
\toolname is sound by design: if a proof is successfully produced, the target VC generated by the VC generator is eventually discharged, with all its dependent helper lemmas formally stated and machine-checked in \rocq. The trusted computing base includes the small \rocq and \framac kernels --- the underlying LLM is not part of the trusted computing base, and the soundness guarantees hold in the face of arbitrary LLM behavior. 

\smallskip
\textit{Failure Analysis.} 
We manually examined all (25=11+7+3+4) VCs, as shown in \autoref{fig:succ_venn}, that are proved by \autorocq but not by \toolname, and classified the root cause as follows:

\begin{itemize}[leftmargin=1em,nosep]
    \item \textit{Import errors} (12/25): importing offline lemmas into the proof context fails due to conflicts such as inconsistent definitions.
    \item \textit{Over-fixation on lemmas} (7/25): \toolname expends too many steps on adapting/proving new lemmas online, whereas \autorocq finds similar, existing lemmas in the context and applies them directly.
    \item \textit{Divergent strategies} (6/25): the additional context confuses the LLM, steering \toolname toward a different proving strategy altogether and leading to early aborts.
\end{itemize}

\smallskip
\textit{Threats to Validity.} 
Potential biases in benchmark selection may threaten the internal validity of our evaluation. 
To mitigate this, we work with two
benchmark sets from prior work covering a range of different verification domains, including both SV-COMP programs and complex real-world C programs, ranging across OS modules, parsers, standard libraries, and algorithms.
The external validity of our results may be limited by the specific configurations and versions of the tools used in our evaluation.
To mitigate this threat, we have used the same version of \rocq (\ie 8.18) and the same configurations (\ie gpt-5.2-2025-12-11 with temperature 0) to ensure a fair comparison.
There is also a potential threat of data leakage, as the benchmarks used in our evaluation may have been seen by the underlying LLM during training, potentially leading to bloated results.
To the best of our knowledge, no ground-truth proofs of either benchmark are publicly available, so we evaluate the risk of data leakage as minimal.

\smallskip
\textit{Limitations and Future Work.}
While \toolname demonstrates significant improvements in proving complex VCs, it has limitations.
First, offline lemmas in our evaluation rarely require substantial proofs, because each lemma typically captures a single utility (\autoref{tab:hl_cat}) and is therefore straightforward to prove. As such, the offline synthesizer may fail to produce useful lemmas for VCs requiring deep domain knowledge or intricate reasoning. Future work could incorporate domain-specific heuristics or richer program context to improve lemma quality.
Second, the online adapter currently depends on existing lemmas and proof assistant feedback to guide refinement, and can be invoked anytime during the proving process as the agent sees fit. This sometimes leads to over-fixation on lemma adaptation, as suggested by the failure analysis. More principled adaptation strategies may warrant further investigation.

%% file: sections/related.tex
\section{Related Work} \label{sec:related}

\textit{Lemma Discovery.}
\toolname is closely related to prior work on automated helper lemma synthesis~\cite{johansson2019lemma} to assist automated theorem proving.
Existing approaches broadly fall into two categories.
\emph{Bottom-up} approaches, also referred to as theory exploration, iteratively conjecture propositions from terms and theorems available in the proving context~\cite{johansson2011conjecture, mccasland2017mathsaid, montano2012scheme, kurashige2024cclemma}.
On the other hand, \emph{top-down} approaches synthesize lemmas by analyzing a given proof state~\cite{sivaraman2022lfind, brendel2025synthesizing}, typically through generalization or rewriting. These tools are often used in conjunction with neural theorem provers to recover from a failed proof.
For both approaches, the conjectures are often constructed from templates and validated through counterexample checking~\cite{denes2014quickchick}. 
To the best of our knowledge, however, existing approaches are limited to mathematical lemmas in practice due to their inability to instantiate counterexamples involving non-primitive types. 
\toolname similarly features a bottom-up phase, through interpretation of both the semantics-aware and proof-targeted VC offline (\autoref{sec:approach:offline_discovery}), and a top-down phase, through on-the-fly refinement to accommodate evolving proof states online (\autoref{sec:approach:online_refinement}).
\toolname synthesizes both the statement and the proof of helper lemmas in \rocq to ensure their validity, instead of searching for counterexamples opportunistically.

\smallskip
\textit{Failure Recovery in Proof Automation.}
Recovering from failed proof attempts is critical for proof automation systems.
Earlier works take binary signals from the proof assistant and retry in case of failures~\cite{sanchez2020generating,sanchez2024qedcartographer,rango-icse25}, whereas others leverage more granular diagnostic messages from the ITP~\cite{lu2025palm, copra24, autorocq}.
These error messages can be helpful for both whole-proof generation~\cite{firstBaldurWholeProofGeneration2023} and heuristic repair of tactics~\cite{luAdaptiveProofRefinement2025,lu2025palm}.
Agentic approaches have introduced more flexibility into the recovery stage by optionally restoring to a prior state~\cite{copra24} or gathering additional context~\cite{autorocq}.
\toolname continues this line of work on leveraging the feedback from the ITP.
However, rather than focusing on failed tactics, \toolname is able to recover at the level of helper lemmas through different means of adaptation. This allows \toolname to refine lemmas curated from other sources (e.g., by the offline synthesizer) and apply them in new proof contexts, significantly reducing the proving efforts.

\smallskip
\textit{Agentic Software Quality Assurance.}
Beyond theorem proving, LLMs have also been applied to upstream stages of the deductive verification pipeline, such as automated inference of specifications~\cite{endres2024can,WANG:OOPSLA2026,specgen-icse25,wen2024enchanting} and assertion hints~\cite{mugnier2025laurel,AutoVerus}. 
More broadly, it has transformed a wide spectrum of software quality assurance techniques~\cite{hou2024large,fan2023large}, \eg model checking~\cite{zuo2025pat,llm-model-checking}, test generation~\cite{deng2024large,xia2024fuzz4all, cottontail-sp26, ConcoLLMic}, and static analysis~\cite{li2024enhancing, wu2024advscanner, guorepoaudit}.
Among these approaches, deductive verification offers unmatched mathematical rigor, making it indispensable for safety-critical systems such as the Linux kernel. 
\toolname advances this frontier by enabling scalable, semantics-driven proof automation that can discharge complex VCs from real-world C programs, addressing a longstanding challenge in the field.

%% file: sections/conclusion.tex
\section{Conclusion} \label{sec:conclusion}

We presented \toolname, a semantic proof agent that synthesizes and adapts helper lemmas for proving verification conditions from real-world C programs. 
\toolname integrates an offline lemma synthesizer, which derives helper lemmas from program semantics, with an online lemma adapter that refines lemmas through feedback-guided adaptation during proof search, thus bridging high-level program semantics and low-level tool-specific details. 
More broadly, \toolname enables semantics-driven proof automation by grounding LLMs' informal reasoning over program semantics to rigorous helper lemmas.
As AI-generated code proliferates, scalable automated verification becomes essential; we envision this paradigm enabling trustworthy deployment of AI-synthesized programs through formal correctness guarantees.

%% file: sections/data-statement.tex
\section*{Disclosure of Generative AI Usage}

LLMs were used in the work as they form a fundamental part of the approach.
In particular, ChatGPT was used as the backend by \toolname and comparison tools in the evaluation. 
ChatGPT were also used by the authors during the development of \toolname, and for grammatical edits of the paper.
The authors validated the results as needed and take full responsibility for the final content.

\section*{Data Availability Statement}

We release \toolname, including its implementation, benchmarks, and replication instructions, for
academic use at the following link:

\noindent
\begin{center}
\url{https://github.com/NUS-Program-Verification/LemmaNet}.
\end{center}

\begin{comment}
\noindent
The artifact is also archived at:

\noindent
\begin{center}
\url{https://doi.org/10.6084/m9.figshare.33014150}.
\end{center}
\end{comment}